\documentclass[fleqn,usenatbib]{mnras}

\usepackage{newtxtext,newtxmath,xcolor}

\usepackage[T1]{fontenc}

\DeclareRobustCommand{\VAN}[3]{#2}
\let\VANthebibliography\thebibliography
\def\thebibliography{\DeclareRobustCommand{\VAN}[3]{##3}\VANthebibliography}

\usepackage{graphicx}	
\usepackage{amsmath}	
\usepackage{stfloats}
\graphicspath{{figure/}}
\usepackage{soul}

\newcommand{\msun}{\mbox{${\rm M}_{\odot}$}}

\newcommand{\mvir}{M_{\rm vir}}
\newcommand{\ct}{{\sc consistent trees}}
\newcommand{\rockstar}{{\sc rockstar}}
\newcommand{\subfind}{{\sc subfind}}

\setstcolor{red}

\def\lesssim{\lower.5ex\hbox{$\; \buildrel < \over \sim \;$}}
\def\gtrsim{\lower.5ex\hbox{$\; \buildrel > \over \sim \;$}}

\title[Santa Cruz SAM compared with IllustrisTNG -- I.]{Galaxy Formation in the Santa Cruz semi-analytic model compared with IllustrisTNG -- I. Galaxy scaling relations, dispersions, and residuals at $z=0$}

\author[Gabrielpillai et al.]{Austen Gabrielpillai$^{1,2,3}$\thanks{a.gabrielpillai@gmail.com},
Rachel S. Somerville$^{4}$,
Shy Genel$^{4,5}$,
Vicente Rodriguez-Gomez$^{6}$,
\newauthor
Viraj Pandya$^{4,7,8,9}$,
L. Y. Aaron Yung$^{2}$,
and Lars Hernquist$^{10}$
\\
$^{1}$Institute for Astrophysics and Computational Sciences, Catholic University of America, USA\\
$^{2}$Astrophysics Science Division, NASA/GSFC, 8800 Greenbelt Rd, Greenbelt, MD 20771, USA\\
$^{3}$Center for Research and Exploration in Space Science and Technology, NASA/GSFC, 8800 Greenbelt Rd, Greenbelt, MD 20771, USA\\
$^{4}$Center for Computational Astrophysics, Flatiron Institute, 162 5th Ave, New York, NY 10010, USA\\
$^{5}$Columbia Astrophysics Laboratory, Columbia University, 550 West 120th Street, New York, NY 10027, USA\\
$^{6}$Instituto de Radioastronom\'ia y Astrof\'isica, Universidad Nacional Aut\'onoma de M\'exico, Apdo. Postal 72-3, 58089 Morelia, Mexico\\
$^{7}$UCO/Lick Observatory, Department of Astronomy and Astrophysics, University of California, Santa Cruz, CA 95064, USA\\
$^{8}$Department of Astronomy, Columbia University, 550 W 120th Street, New York, NY 10027, USA\\
$^{9}$Hubble Fellow\\
$^{10}$Center for Astrophysics | Harvard \& Smithsonian, 60 Garden St., Cambridge, MA 02138, USA\\
}

\date{Accepted XXX. Received YYY; in original form ZZZ}

\pubyear{2022}

\begin{document}
\label{firstpage}
\pagerange{\pageref{firstpage}--\pageref{lastpage}}
\maketitle

\begin{abstract}
We present the first results from applying the Santa Cruz semi-analytic model (SAM) for galaxy formation on merger trees extracted from a dark matter only version of the IllustrisTNG (TNG) simulations. 
We carry out a statistical comparison between the predictions of the Santa Cruz SAM and TNG for a subset of central galaxy properties at $z=0$, with a focus on stellar mass, cold and hot gas mass, star formation rate (SFR), and black hole (BH) mass. 
We find fairly good agreement between the mean predictions of the two methods for stellar mass functions and the stellar mass vs. halo mass (SMHM) relation, and qualitatively good agreement between the SFR or cold gas mass vs. stellar mass relation and quenched fraction as a function of stellar mass.
There are greater differences between the predictions for hot (circumgalactic) gas mass and BH mass as a function of halo mass. Going beyond the mean relations, we also compare the dispersion in the predicted scaling relations, and the correlation in residuals on a halo-by-halo basis between halo mass and galaxy property scaling relations. Intriguingly, we find similar correlations between residuals in SMHM in the SAM and in TNG, suggesting that these relations may be shaped by similar physical processes. Other scaling relations do not show significant correlations in the residuals, indicating that the physics implementations in the SAM and TNG are significantly different. 

\end{abstract}




\section{Introduction}

In the highly successful standard $\Lambda$ Cold Dark Matter model ($\Lambda$CDM) of cosmological structure formation, small initial density fluctuations are amplified by gravity in the expanding Universe, and gravitationally bound dark matter dominated objects (dark matter halos) form when their local gravity overcomes the effect of cosmic expansion. These halos form the sites where gas can cool and collapse to high densities, eventually forming stars \citep{White&Rees:1978, white-frenk:1991}. There is general consensus that, within this backbone of dark matter halo collapse and growth, a broad suite of intertwined physical processes operating over a vast range of scales interact to shape observable galaxy properties \citep{SD15,naab:2017}. In addition to gravity, processes thought to be important include radiative cooling of gas, formation of stars out of dense gas, feedback from massive stars and supernovae, and the formation, growth, and feedback from supermassive black holes. In order to interpret the wealth of available and upcoming multi-wavelength galaxy observations, it is important to develop theoretical models that can capture the essential physics that shapes galaxy properties. Moreover, it is critical to exploit observations that can probe \emph{multiple components} of galaxies, such as stellar mass, cold interstellar medium (ISM) gas mass, warm/hot circumgalactic medium (CGM) mass, metal mass in different components, star formation rate, etc. 

Two broad classes of modeling techniques have been developed and widely used to make self-consistent physics-based predictions of galaxy properties in the context of a $\Lambda$CDM cosmology \citep{SD15}. Semi-analytic models (SAMs) \citep{white-frenk:1991, Cole:1994, Kauffman:1994, Somerville:1999}, are set within the skeleton of a suite of merger trees, which may be extracted from dark matter only N-body simulations or constructed via semi-analytic methods based on the Press-Schechter formalism \citep[see][and references therein]{Lacy&Cole:1993,SK:99}. In the absence of feedback processes, gas accretion into halos is assumed to track the accretion of dark matter. Empirical or phenomenological recipes are used to model the radiative cooling of gas from the hot CGM into the ISM, the conversion of cold ISM gas into stars, and the rate that gas is heated and ejected by stellar feedback. Similar recipes are used to track chemical evolution, and to model black hole (BH) seed formation, BH growth via accretion and mergers, and heating and winds driven by this BH accretion, as well as other processes which vary from one model to another. SAMs are fundamentally a type of ``flow model'', in that they consist of solving ordinary differential equations governing the flow of material between different reservoirs (intergalactic gas, CGM, ISM, stars, etc). Many SAMs track only global galaxy properties (i.e., the total mass within these different reservoirs), and do not provide detailed information on the spatial distribution of baryons, but some have been extended to provide basic estimates of the size and internal structure of galaxies \citep[e.g.][]{somerville:2008b,porter:2014}. When SAMs are run within halos extracted from N-body simulations, they can be used to make predictions for galaxy clustering \citep[e.g.][]{kauffmann:1999, somerville:2001, boryana:2021}. 

Another powerful technique to model galaxy formation is numerical simulations, which explicitly solve the equations of gravity, (magneto)hydrodynamics, and thermodynamics, for discretized particles or grid cells representing dark matter, gas, stars, and black holes \citep{SD15,naab:2017}. As a result, numerical simulations provide detailed information on the spatial distribution, temperature, and kinematics of all of these components as galaxies form and evolve over cosmic time. However, many of the critical processes involved in galaxy formation cannot be resolved and directly simulated in current cosmological simulations, including star formation, stellar feedback, and black hole formation, growth, and feedback. As a result, these processes are implemented in numerical simulations using ``sub-grid recipes'', which in many cases are phenomenological and quite closely resemble those implemented in SAMs. For example, the mass loading of stellar driven winds may be assumed to be a function of the halo velocity dispersion, and put in ``by hand''. In other cases, pragmatic mitigations such as switching off cooling or hydrodynamic forces, or storing up energy and releasing it in a ``thermal bomb'', are introduced, in an attempt to represent processes that are not resolved or that are not explicitly included in the simulation. 

Both techniques, therefore, contain parameters that characterize the phenomenological sub-grid recipes. Since the underlying physics is often poorly understood, it has become common practice to tune or ``calibrate'' these parameters to reproduce a subset of galaxy observations. The choice of which observations are used to calibrate the model parameters, as well as the required level of precision in reproducing the observations, varies from one study to another, introducing an additional source of uncertainty in comparing and interpreting the results. 

Once calibrated in this way, both techniques have shown comparable success in reproducing and predicting a wide variety of galaxy observables over cosmic time \citep[see e.g.][]{SD15}. Overall, the two techniques (semi-analytic modeling and numerical simulations) have also yielded similar paradigms in terms of which physical processes are important in shaping galaxy properties as a function of scale; for example, in both types of models, stellar feedback makes star formation inefficient in low-mass halos, while feedback from Active Galactic Nuclei (AGN) quench cooling flows and star formation in massive halos. Clearly, one of the main differences between the techniques is the computational expense: the computational requirements for SAMs are many orders of magnitude smaller than those for numerical hydrodynamic simulations. On the other hand, SAMs must incorporate assumptions and simplifications with an unknown degree of accuracy. 

There have been numerous studies comparing the predicted galaxy properties in a particular numerical hydrodynamic simulation with a particular semi-analytic model \citep{Benson:2001,yoshida:2002,helly:2003,cattaneo:2007,stringer:2010,hirschmann:2012,guo:2016,popping:2019,ayromlou:2021}. Some studies carried out statistical comparisons of the populations in both types of models, while others have run the SAM within dark matter halo merger trees extracted from the N-body simulation (or a dark matter only version run with the same initial conditions) and carried out both halo-by-halo and galaxy-by-galaxy comparisons. 

However, an important question that remains unanswered is whether the two techniques are mapping from dark matter formation history to observable galaxy properties in \emph{a fundamentally similar way}, or whether the agreement is due to the similar calibration, or degeneracies in physical processes and the commonly studied observables. For example, \citet{Pandya:2020} recently compared results from the FIRE-2 numerical hydrodynamic simulations \citep{fire2} with those of the Santa Cruz SAM (SC-SAM) for a suite of halos with masses ranging from $10^{10}$--$10^{12}\, \msun$. They found very good agreement between FIRE-2 and the SC-SAM for stellar masses and cold ISM mass, but substantial disagreement in the properties of the hot halo gas (circumgalactic medium) --- a quantity that has not been widely used up to now to calibrate or validate models, at least for Milky Way and lower mass halos. \citet{Pandya:2020} further found, when they analyzed the flow cycle for these halos, that particularly in the lower mass halos, the two models achieved similar outcomes with regard to the stellar and ISM mass by very different means --- the SAMs had much larger inflow rates, which were compensated by higher outflow rates. 

In this work, we compare the predictions of the TNG numerical hydrodynamic simulations \citep{pillepich:2018,nelson:2018,2018MNRAS.477.1206N,2018MNRAS.480.5113M,springel:2018} with those of the Santa Cruz SAM \citep{Somerville:1999,somerville:2008,somerville:2015}. As SAMs can be sensitive to the details of the merger tree algorithm, we run the \rockstar\ halo finding code \citep{rockstar} and the \ct\ merger tree builder \citep{behroozi:ct} on the dark matter only versions of TNG. The Santa Cruz SAM has been extensively tested and calibrated using these halo/merger tree methods run on the Bolshoi series of N-body simulations \citep{porter:2014,Lu:2014}. We create bijective matches between the \rockstar\ halos and the native TNG subfind halos, so that we can do a halo-by-halo comparison. We expand on other recent comparisons of TNG with SAMs \citep[e.g.][]{ayromlou:2021} in that in addition to comparing galaxy scaling relations stastically and halo by halo, we compare the \emph{dispersion} in the predicted scaling relations at fixed halo (and stellar) mass, and we also compare the correlations between \emph{residuals} of halos from their respective scaling relations for matched halos in TNG and the SAM. To our knowledge, this type of residual-residual analysis comparison between numerical simulations and SAMs has not been carried out in a systematic way before. In this work, we focus on galaxies identified at $z=0$; in Paper II, we present the corresponding analysis and results for galaxy formation histories over cosmic time and comparisons at different snapshots in cosmic time. Although there is a much larger set of predicted properties that we could compare in this way, we choose in this paper to focus on stellar mass, cold gas mass, hot gas (CGM) mass, star formation rate (SFR), and black hole mass. In order to simplify the analysis, we also choose to focus on \emph{central} galaxies only. This is because the Santa Cruz SAM does not make use of the sub-structure information from the N-body simulation, instead tracking satellite evolution using an internal semi-analytic treatment, and therefore it is not possible to conduct a halo-by-halo comparison for satellites.

This work represents one of the first steps in a longer term program, the goal of which is to determine whether traditional SAMs can be modified to ``emulate'' the predictions of numerical hydrodynamic simulations with enough fidelity that they can be used to extend the reach of numerical techniques. For example, SAMs can be used to do automated parameter space exploration and Bayesian inference with techniques such as
Markov Chain Monte Carlo
(MCMC; e.g. \citealp{Lu:2011}; \citealp{Henriques:2015}) or Simulation Based Inference \citep[e.g.][]{alsing:2019}, which require the simulations to be run many thousands (to tens of thousands) of times. In addition, SAMs can be used to create simulations with higher dynamic range (i.e. simulating galaxies down to lower masses and/or spanning larger volumes) than full numerical hydrodynamics. In order to make forecasts for and interpret results from the upcoming generation of large galaxy surveys (e.g. DESI, Vera Rubin Observatory, Euclid, and the Nancy Grace Roman Space Telescope), and Line Intensity Mapping experiments (see \citealp{kovetz:2017} for a review), it is clear that new approaches that can dramatically expand the dynamic range of theoretical, physics-based simulations will be needed. 

The structure of the paper is as follows. In \S\ref{sec:models}, we provide a brief background on the main ingredients of the TNG simulations and the Santa Cruz semi-analytic models. In \S\ref{sec:methods}, we provide details on the halo finding and merger tree algorithms used in this work. In \S\ref{sec:results}, we present the comparison of galaxy property distributions, scaling relations, and residuals.  We discuss our results in \S\ref{sec:discussion} and summarize and conclude in \S\ref{sec:conclusions}. Throughout our analysis, we adopt values for the cosmological parameters that are consistent with the constraints obtained by the Planck Collaboration \citep{Planck2015} for calibrating the SAM; these are the same as those used in the TNG runs ($\Omega_{m}$ = 0.3089, $\Omega_{\Lambda}$ = 0.6911, $\Omega_{b}$ = 0.0486, and $h$ = 0.6774). 

\section{Models}
\label{sec:models}
In this section, we briefly outline the two models used in this work -- the TNG suite and the Santa Cruz semi-analytic model. We focus on the physical processes that are most relevant to the quantities we have chosen for this analysis (stellar mass, cold gas mass, hot gas (CGM) mass, SFR, and black hole mass).  We refer to the works mentioned in each sub-section for a full description of the models.

\subsection{IllustrisTNG hydrodynamic simulations}
\label{sec:models:TNG}
The Next Generation Illustris simulations (IllustrisTNG; \citealp{springel:2018,weinberger:2017,pillepich_TNGmethods:2018,nelson:2018}) are a suite of cosmological magneto-hydrodynamical simulations based on the moving-mesh refinement code AREPO \citep{Springel:2010, Pakmor:2011, Pakmor:2016}, 
developed to understand galaxy formation and evolution in a large-scale environment. In TNG, several of the physics modules were modified relative to its predecessor, Illustris \citep{vogelsberger:2014, vogelsberger:2014b, genel:2014}, to produce improved agreement with key observations. The main modifications were to the treatment of AGN feedback \citep{weinberger:2017}, and to stellar driven winds \citep{pillepich_TNGmethods:2018}. In addition, some modifications were made to the treatment of magnetic fields, and improvements to the flexibility and hydrodynamic convergence of the code were made. There are three main simulation sets that comprise the TNG suite: TNG50 \citep{pillepich:2019,Nelson:2019}, TNG100, and TNG300 \citep{TNGdatarelease}. Each set adopts a different box size and contains a series of runs that have varying mass resolutions. Each run has a dark-matter only (DMO) and an analogous full physics (FP) run with the same initial conditions. See Table \ref{tab:tng_table} for details on the simulation sets used in this work.

\begin{table}
	\centering
	\begin{tabular}{lrrr} 
		\hline
		Simulation & $L_{\rm Box} [{\rm cMpc} / h]$ & $N_{\rm DM}$ & $m_{\rm DM} [\msun / h]$  \\
		\hline
		TNG100-1      & 75  & $1820^{3}$ & $5.06 \times 10^{6}$ \\
		TNG100-1-Dark & 75  & $1820^{3}$ & $6.00 \times 10^{6}$ \\
		TNG300-1      & 205 & $2500^{3}$ & $3.98 \times 10^{7}$ \\
		TNG300-1-Dark & 205 & $2500^{3}$ & $4.73 \times 10^{7}$ \\
		\hline
	\end{tabular}
    \caption{Summary of TNG boxes used in this study. $L_{\rm Box}$ is the box side length, $N_{\rm DM}$ is the number of dark matter particles, and $m_{\rm DM}$ is the mass of a dark matter particle.}  
	\label{tab:tng_table}
\end{table}

Processes that are treated via ``sub-grid'' recipes in TNG include star formation, stellar feedback, black hole seeding, black hole accretion, and AGN feedback. We very briefly summarize these prescriptions here, and refer to \citet{pillepich_TNGmethods:2018} and references therein for details. Following the sub-grid physics implementation in the original Illustris simulation \citep{VogelsbergerM_13a}, the model of \citet{SH:2003} is adopted to treat star formation and the pressurization of the interstellar medium (ISM).  Stellar driven winds are modeled by injecting kinetic energy into gas cells with a probability specified by a parameterized mass loading function \citep{pillepich_TNGmethods:2018}. In TNG, this function depends on the gas phase metallicity and the dark matter (DM) velocity dispersion $\sigma_{\rm DM}$, measured with a weighted kernel over the nearest 64 DM particles. The velocity imparted to the wind particles is also a function of $\sigma_{\rm DM}$ and redshift. Every halo above a critical mass receives a seed black hole with a fixed mass. The rate that gas can accrete onto nuclear BHs is given by a standard Eddington-limited Bondi-Hoyle model. Highly energetic BH-driven winds are launched when the BH accretion rate is below a critical Eddington rate and the BH mass is above a critical value, again by injecting kinetic energy into neighboring gas cells \citep{weinberger:2017}. 

Each of the sub-grid processes in TNG contains one or more adjustable parameters. These parameters are calibrated to approximately reproduce a selected set of observations. The observations used for calibration of TNG are discussed in \citet{pillepich_TNGmethods:2018}, and include the stellar mass function, cosmic SFR density as a function of redshift, BH mass vs. stellar mass relation at $z=0$, hot gas fraction in galaxy clusters at $z=0$, and the galaxy stellar mass vs. radius relation at $z=0$. 

Gas in the ISM is not explicitly partitioned into molecular, atomic, and ionized components on the fly in TNG. However, these quantities have been calculated in post-processing by \citet{diemer:2018} and \citet{diemer:2019}, further examined in\citet{Stevens:2021}, and made available as a supplemental data catalog on the TNG data portal. For compatibility with the gas partitioning approach used in the Santa Cruz SAM (described below), we use the catalogs that adopt the \citet{GK11} model (GK11) and the ``volumetric'' method, where the molecular fraction is computed cell by cell, and the surface density is obtained by multiplying by the Jeans length. For more details and a comparison of different methods for computing the gas partioning in IllustrisTNG, please see \citet{diemer:2018} and \cite{diemer:2019}. These catalogs are complete above a minimum stellar or gas mass of $2 \times 10^{8} \msun$.

\subsection{Santa Cruz Semi-Analytic Model}
\label{sec:models:SAM}

The version of the Santa Cruz SAM used here is very similar to the one published in \citet{somerville:2015}, and used in other recent papers such as \citet{Yung:2019a,Yung:2019b} and \citet{somerville:2021}. We refer the reader to these papers, along with \citet{somerville:2008} and \citet{porter:2014} for details. 

The backbone of the Santa Cruz SAM is a suite of dark matter halo merger trees, which are described in more detail in \S\ref{sec:methods:halosandtrees}. The merger tree provides information on the mass of collapsed dark matter halos at a given snapshot, and which of these halos merge together at a subsequent timestep. They also provide information on the growth of halos by accretion of ``diffuse'' material (particles that are not in halos) over time. 

When a halo enters the virial radius of another, larger halo, it becomes a ``sub-halo''. Galaxies that are hosted by sub-halos are called satellites, and the galaxy that resides at the center of the main host halo is called the central. The Santa Cruz SAM does not use the information on sub-halo positions or disruption status from the N-body simulation, and instead uses a semi-analytic model to track the orbital decay and tidal destruction of sub-halos and the satellite galaxies that they host (see \citealt{somerville:2008} (S08) for details).  

In the absence of any feedback processes, the rate that gas flows into a halo is given by $f_b \dot{M}_h$, where $f_b$ is the universal baryon fraction and $\dot{M}_h$ is the growth rate of the dark matter halo. After the Universe is reionized, gas inflow into halos is suppressed due to the meta-galactic photoionizing background. We assume that the Universe is fully reionized by $z=8$, and use the filtering mass based on results from numerical hydrodynamic simulations by \citet{Okamoto:2008}. The fraction of baryons that is able to accrete into the halo as a function of halo mass is given by Eqn. 3 in S08.

Gas that accretes into the halo forms a ``hot halo'' or circumgalactic medium (CGM). This gas is assumed to be isothermal and to have a spherical isothermal density profile given by $\rho \propto r^{-2}$. The ``cooling radius'' $r_{\rm cool}$ is computed as the radius within which gas at temperature $T_{\rm vir}$ and metallicity $Z_{\rm hot}$ has had time to radiate away all of its energy, using standard radiative cooling functions from \citet{SD:1993}. The rate that gas cools and is accreted into the ``cold gas reservoir'' (ISM) when $r_{\rm cool}$ is less than the virial radius $R_{\rm vir}$ is given by Eqn. 2 in S08. When $r_{\rm cool} > R_{\rm vir}$ (corresponding to a cooling time that is less than the dynamical time), it is assumed that the cooling rate is given by the rate that gas accretes into the hot halo. 

We assume that the cold ISM gas forms a disk with a radial exponential profile, which we compute using the assumption that specific angular momentum is conserved, and that the baryons cause adiabatic contraction of the halo when they fall in (see \citealt{somerville:2008b} for details). Based on this gas density profile, we compute the fraction of the cold gas in each radial annulus that is in the form of ionized, atomic, and molecular gas (${\rm H}_2$) using the approach outlined in detail in \citet{somerville:2015}. The star formation rate is then computed by assuming a scaling relation between the molecular gas surface density and star formation rate surface density, as motivated by observations. The version of the SC SAM used here adopts the GK11 recipe for gas partitioning, in which the ${\rm H}_2$ fraction depends on gas surface density, gas phase metallicity, and the background radiation field (assumed to be proportional to the local SFR), based on fitting formulae extracted from the numerical hydrodynamic simulations of \citet{GK11}. The star formation recipe adopted here is the ``Big2'' recipe \citet{Big2}, in which the slope of the $\Sigma_{\rm SFR}$ vs $\Sigma_{\rm H_2}$ relation steepens above a critical surface density \citep[see][for details]{somerville:2015}. 

Stellar feedback ejects cold gas from the ISM with a rate given by 
\begin{equation}
\dot{m}_{\rm eject} = \epsilon_{\rm SN} \left(\frac{200\, \rm{km/s}}{V_{\rm disk}}\right)^{\alpha_{\rm rh}} \dot{m}_*
\end{equation}
where $\epsilon_{\rm SN}$ and $\alpha_{\rm rh}$ are adjustable parameters, $V_{\rm disk}$ is the circular velocity of the disk, approximated as the circular velocity of the (uncontracted) halo at twice the Navarro-Frenk-White (NFW) scale radius $r_s$ \citep{NFW}, and $\dot{m}_*$ is the star formation rate. A fraction $f_{\rm eject}$ of this gas is ejected from the halo and deposited in an ``ejected'' reservoir, while the rest is assumed to be heated to the virial temperature, and is deposited in the hot gas halo (CGM).  The function $f_{\rm eject}$ is a simple power law function of halo circular velocity (see Eqn. 13 in S08). Gas in the ejected reservoir is ``re-accreted'' into the hot halo with a rate $\chi_{\rm re-infall} \, m_{\rm eject}/ t_{\rm dyn}$, where $\chi_{\rm re-infall}$ is an adjustable parameter, $m_{\rm eject}$ is the mass of gas in the ejected reservoir, and $t_{\rm dyn}$ is the dynamical time of the halo.

Each top-level DM halo (i.e., halo with no progenitors) is seeded with a black hole with a mass $m_{\rm BH, seed}$. The black hole can accrete mass through two modes, one fed by cooling flows from the hot halo (see Eqn. 20 in S08), and one by inflows of cold gas from the ISM driven by mergers or internal gravitational instabilities (see Section 2.9 in S08 and Section 3.3 in \citet{hirschmann_bh:2012}). Cold gas can be ejected from the ISM by the merger or disk instability driven ``radiative mode'' AGN-driven winds (see Section 2.10 of S08). In addition, the cooling flow mode is assumed to produce radiatively inefficient accretion onto the BH that gives rise to radio jets, which can heat the hot halo (CGM) gas (see Section 2.11 and Eqn. 21 of S08). 

\begin{table}
	\centering
	\caption{Summary of selected SC SAM parameters}
	\label{tab:param}
	\begin{tabular}{ l  l  c } 
		\hline
		Parameter       & Description              & Value \\ [0.5ex] 
		\hline
		$\epsilon_\text{SN}$ &  SN feedback efficiency  & 1.7   \\ 
		$\alpha_\text{rh}$ & SN feedback slope & 3.0 \\
                $V_{\rm eject}$ & halo gas ejection scale &  110 km/s \\
                $\chi_{\rm re-infall}$ & re-accretion timescale for ejected gas & 0.1 \\
		$\tau_{*,0}$ &  SF timescale normalization & 1.0   \\ 
		$y$ & Chemical yield (in solar units) & 1.2 \\
		$\kappa_\text{AGN}$ & Radio mode AGN feedback & $2.0$ $ \times 10^{-3}$ \\
		$m_{\rm BH, seed}$ & mass of seed BH & $10^4 \msun$ \\
		\hline
	\end{tabular}
\end{table}

\section{Methods: Halo finders, Merger Trees, and Bijective Matches}
\label{sec:methods}
The halo finder and merger tree algorithm provide the fundamental framework for SAMs. The Santa Cruz SAM has been extensively tested and validated using the \rockstar\ halo finder \citep{rockstar} and the \ct\ merger tree algorithm \citep{behroozi:ct}. For TNG, a different halo finder based on the friends-of-friends \citep{FoF} (FoF) algorithm is used to identify ``groups'', and the \subfind\ algorithm is used to identify substructure. The carefully measured baryonic properties made available by the TNG team are defined using these halo/subhalo catalogs, and we make use of these. The merger trees in TNG are constructed with the SubLink algorithm \citep{Rodriguez-Gomez:2015}. In unpublished work, the SC SAM group has found that the predictions of the SC SAM are very sensitive to the number of mergers identified by the merger tree algorithm, and that the SAM does not perform well when run within merger trees based on SubLink. Therefore, we have run \rockstar\ and \ct\ on the dark matter only versions of the TNG simulations, and then created bijective matches between the \rockstar\ and FoF/\subfind\ halos. In this section we briefly summarize the halo finder and merger tree algorithms, and describe the method used to create bijective matches between the two different sets of halo catalogs. 

\subsection{\rockstar \& \ct}
\label{sec:methods:halosandtrees}
\subsubsection{The {\sc rockstar} halo finder}

\rockstar\ is a phase-space based halo finder developed by \citet{rockstar}\footnote{\url{https://bitbucket.org/gfcstanford/rockstar}}. As an initial step, particles are grouped into 3D friends-of-friends groups with a linking length of $b=0.28$, given in units of mean inter-particle separation, which is somewhat larger than that expected to correspond to traditional halo virial radii. This step allows for easy parallelization, while ensuring that the FoF groups encompass even the most ellipsoidal halos. Next, the algorithm builds a hierarchy of FoF groups in 6D phase space by tuning the linking length so that a specified fraction of the particles in each parent group are captured in each subgroup. Then, this hierarchy of groups is converted into particle membership for halos. Host halo/sub-halo relationships are computed using information from the previous simulation snapshot. Finally, unbound particles are removed, and halo properties are computed. 

For halo masses, \rockstar\ computes the mass within a spherical overdensity that is specified by the user. Many different conventions exist for the value of this overdensity; by default \rockstar\ stores several of the most commonly used ones. All results presented here make use of the halo virial mass definition of \citet{bryan:1998}, given in Eqn.~1 of \citet{2016MNRAS.462..893R}. \rockstar\ also carries out a fit of the radial profile to a \citet{NFW} functional form and records the best fit concentration parameter $c_{\rm NFW}$ \citep[see][for details]{rockstar} and the dimensionless spin parameter, both of which are used by the SAM. 

As SAMs are traditionally run in merger trees extracted from dark matter only simulations, we ran \rockstar\ and consistent trees on the dark matter-only versions of the TNG simulations. We constructed the halo catalogs and merger trees for the highest resolution runs of two TNG volumes: TNG100-1-Dark, TNG300-1-Dark. 
To validate our results, we compared the halo mass functions and other properties with those in the publicly available \rockstar\ catalogs based on the Bolshoi-Planck simulation (\citealt{2016MNRAS.462..893R}). We also performed comparisons and sanity checks between the \rockstar\ catalogs and the TNG team FoF/\subfind\ catalogs, as described below. 

\subsubsection{The \ct\ merger tree builder}

\ct\ \citep{behroozi:ct}\footnote{\url{https://bitbucket.org/pbehroozi/consistent-trees/src/main/}} is a merger tree builder that was designed to ensure gravitational consistency of halo properties across timesteps. Halo descendants are first identified using a traditional particle based algorithm (i.e., being a descendant requires that a halo in a subsequent timestep contains a certain fraction of the particles from the progenitor). Then, the descendant positions and velocities are evolved back in time to estimate their most likely positions at the previous timestep. This step allows the branch to be cut for spurious descendants. A search is performed over several timesteps to identify branches that are ``broken'', i.e. where the progenitor/descendant at a particular timestep is not identified, perhaps due to a merger or other numerical effects. In this case, \ct\ adds in a halo to bridge the gap. \ct\ has demonstrated improvements in both the purity and completeness of merger trees relative to other algorithms. 

We ran \ct\ with the fiducial parameters specified in \citet{behroozi:ct} for the \rockstar\ catalogs from TNG100-1-Dark and TNG300-1-Dark.
Before running the SAMs, we post-process the \ct\ catalogs to remove subhalo trees, as the SC SAM treats substructure internally. We validated the results by visually examining the merger histories, and by comparing the SC SAM results from running on the Bolshoi-Planck merger trees with those that resulted from running them in the new TNG-Dark based merger trees. The results were nearly identical. 

\subsection{\subfind\ halo finder}

\subfind\ \citep{springel:2001} first identifies parent groups using a friends-of-friends algorithm with a standard linking length of $b=0.2$, given in units of mean inter-particle separation. Then, gravitationally bound subhalos within each FoF group are identified by estimating the local density using adaptive kernel interpolation with a specified number of smoothing neighbors. Locally overdense regions are identified as those that are enclosed by an isodensity contour that traverses a saddle point within the density distribution of the candidate halo, and each overdensity is considered a candidate subhalo. Starting from the particle with the highest density, and proceeding to additional particles in decreasing density order, particles are grouped into subhalos. Unbound particles are then removed. Each particle can be a member of only one subhalo. The particle at the minimum of the gravitational potential is adopted as the subhalo center, and the mass is assigned as the sum of the masses of the associated particles. For the main subhalo (host halo), \subfind\ also computes a spherical overdensity based mass.  

We compared the halo mass functions from \rockstar\ at various redshifts with those from \subfind\ applied to both the dark matter only and full physics run of TNG. We find very small differences between the dark matter only (DMO) results from \rockstar\ and \subfind, consistent with the results presented by \citet{gomez:2021} and \citet{boryana:2021}. As a supplementary verification step, we plotted the positions of the most massive halos in the \subfind\ and \rockstar\ DMO catalogs 
along with overlaid circles of radius $r_{\rm vir}$. We found consistency in the positions and radii of massive halos in the two catalogs. 

\subsection{\rockstar-\subfind\ bijective matches}
In order to identify matching halos from the \rockstar\ catalogs with those in the \subfind\ catalogs, we use a software tool called SubLink \citep{Rodriguez-Gomez:2015}\footnote{\url{https://bitbucket.org/vrodgom/sublink}}. 
The use of SubLink for matching catalogs is described in Section 2.3 of \citet{Rodriguez-Gomez:2017}. In particular, eq. (1) defines the merit function for matching subhalos. In this work, we used $\alpha=0$, which reduces the merit function to the trivial case of choosing the subhalo that has the most particles in common.

\begin{figure}
	\includegraphics[width=\columnwidth]{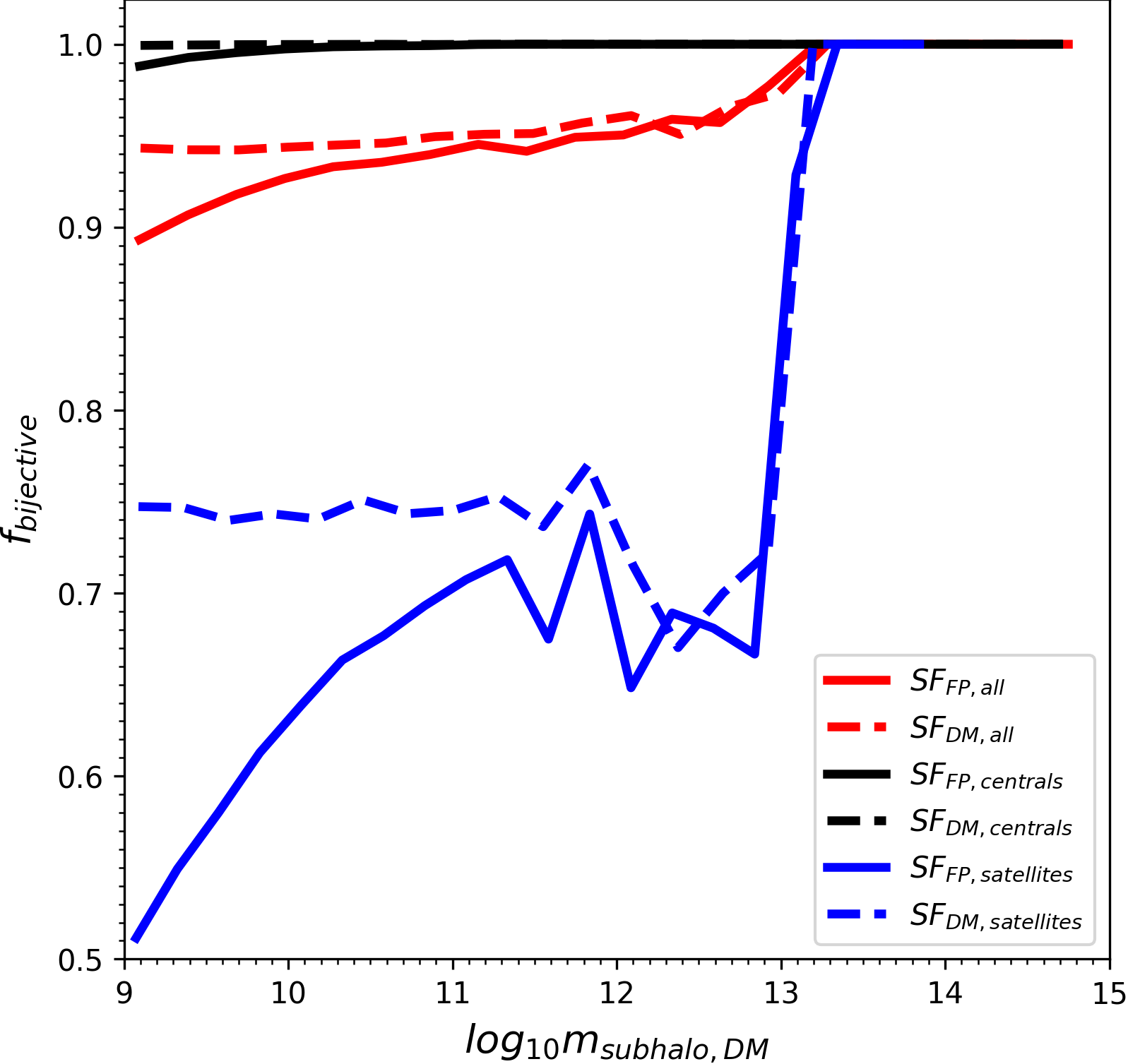}
    \caption{A comparison of the bijective match fraction as a function of subhalo dark matter mass for \subfind-FP (solid) and \subfind-DM (dashed). We compare this fraction for all subhalos (red), as well as just central (black) and satellite (blue) subhalos.
   }
    \label{fig:match-perc}
\end{figure}

As preparation for running SubLink on our new catalogs, we obtained the particle IDs that belong to each subhalo and arranged them in a SubLink-readable format. We utilized an option in the \rockstar\ parameter file to output particle IDs that belong to their respective halos and subhalos. SubLink matches objects from one catalog to another, i.e., it can record the objects in the \rockstar\ catalog that match objects defined by the \subfind\ catalog, or vice versa. We therefore ran the SubLink algorithm in both directions, \subfind-\rockstar\ and \rockstar-\subfind. Objects that are matched in both directions are considered bijective matches. Note that we only attempted to match central galaxies in this study. We carried out bijective matches both between the \rockstar-DMO and \subfind-DMO catalogs, and the \rockstar-DMO and \subfind-FP catalogs. 
The fraction of halos with bijective matches as a function of halo mass at $z = 0$ is shown in Fig.~\ref{fig:match-perc} for both \subfind-FP and \subfind-DMO for the TNG100 and TNG100-Dark runs, for well resolved halos ($>100$ DM particles). We observe that $>99$ \% of centrals have bijective matches. The fraction of bijective matches for non-central subhalos can be considerably lower, from 50--70 percent. 

\begin{figure}
	\includegraphics[width=\columnwidth]{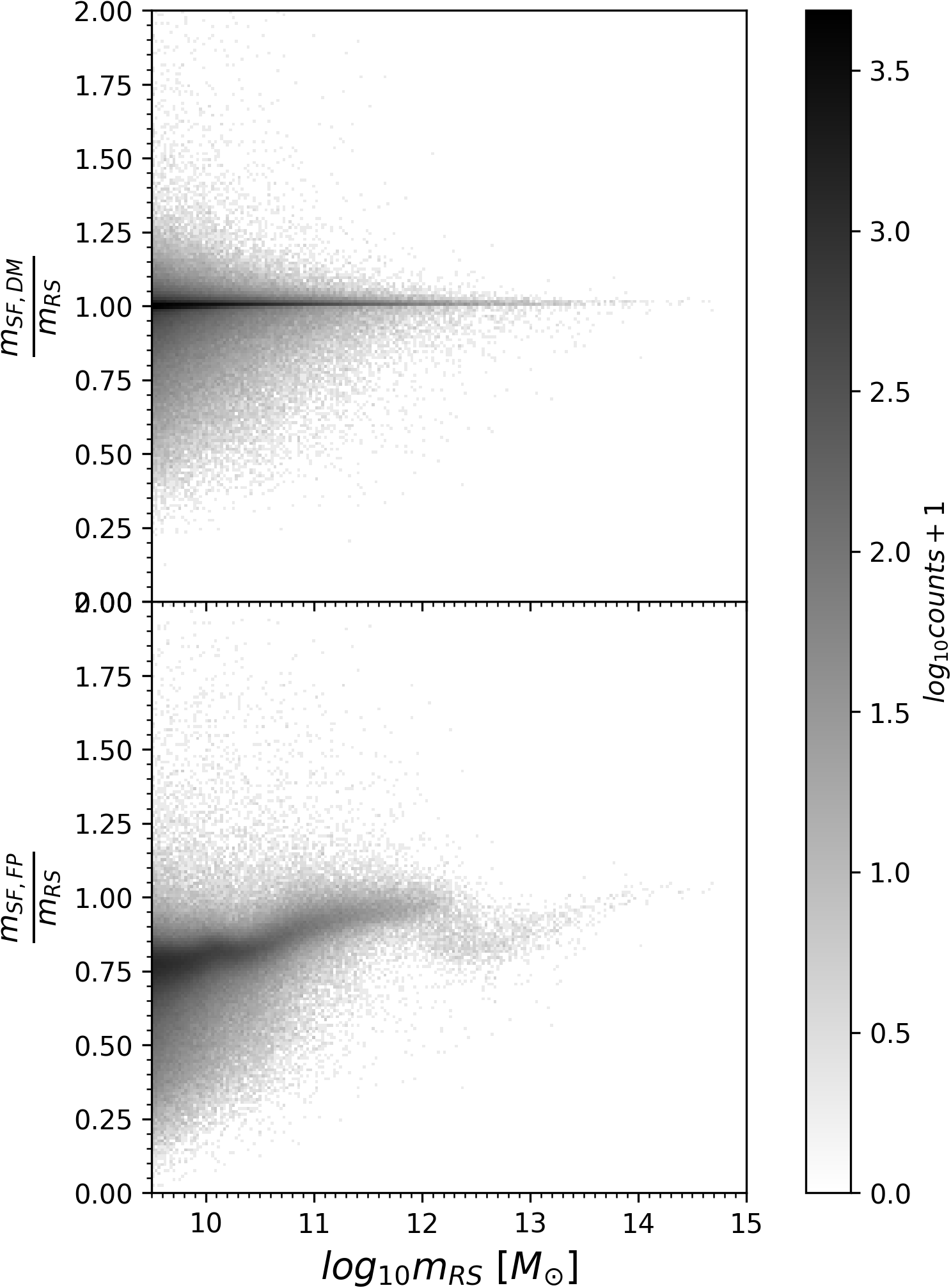}
    \caption{Top panel: log of the ratio between the virial mass for bijectively matched halos in the \subfind\ DMO catalogs and the \rockstar\ DMO catalogs as a function of virial mass. Bottom panel: same as the top panel, but for the \subfind\ full physics catalogs and the \rockstar\ DMO catalogs. The agreement between the DMO catalogs is quite good, while there are larger differences between the DMO and full physics catalogs, which is to be expected due to the effects of baryonic processes on halo properties. The bottom panel is consistent with what is seen in Fig. 16 of \citet{springel:2018}.
   }
    \label{fig:mvir-diff}
\end{figure}

For both of these comparisons, we investigated the difference in $\mvir$, $r_{\rm vir}$, 3D position and 3D velocity between the matched objects as a function of $m_{\rm vir}$.  A comparison of the values of $\mvir$ can be seen in Fig.~\ref{fig:mvir-diff}, where we compare the difference in halo virial mass between the two catalogs. The top panel shows the comparison between \rockstar\ and \subfind\ when both are applied to the DMO catalogs. There is very good agreement, with no significant systematic offset, and a scatter of less than $\pm 0.05$ dex when observed in log-log space. The 3D position and velocity matches also show excellent agreement (see also \citealt{boryana:2021}, who show the clustering properties of halos in these two sets of catalogs). For the comparison between \rockstar-DMO and \subfind-FP (bottom panel), we see a more substantial offset with a trend with halo mass. This offset is also seen when using the matches between \subfind-DMO and \subfind-FP created by LHaloTree \citep{LHaloTree}, showing that these offsets are predominantly due to the effects of baryonic physics and not differences in the halo finding algorithms. 

\section{Results}
\label{sec:results}

There are many physical quantities that are predicted by both the SAM and TNG and could be compared. In this study, we focus on a selected set of properties that quantify star formation, galaxy assembly and quenching: stellar mass ($m_*$), cold neutral ISM mass (here defined as $m_{\rm cold} \equiv m_{\rm HI} + m_{\rm H2}$), star formation rate (SFR), hot gas (circumgalactic medium) mass $m_{\rm hot}$, and black hole mass ($m_{\rm BH}$). Please refer to Appendix~\ref{sec:quantities} for a detailed description of how these quantities are measured and defined in both the TNG simulations and the SAMs. In addition, we need to take into account the differences that arise from the different ways that resolution effects manifest in numerical hydrodynamic simulations versus SAMs. We define a minimum halo mass, stellar mass, gas mass, and SFR for our analysis which are discussed in Appendix~\ref{sec:quantities} and summarized in Table~\ref{tab:fields}.

\subsection{SAM Calibration}

\begin{figure*}
	\includegraphics[width=\textwidth]{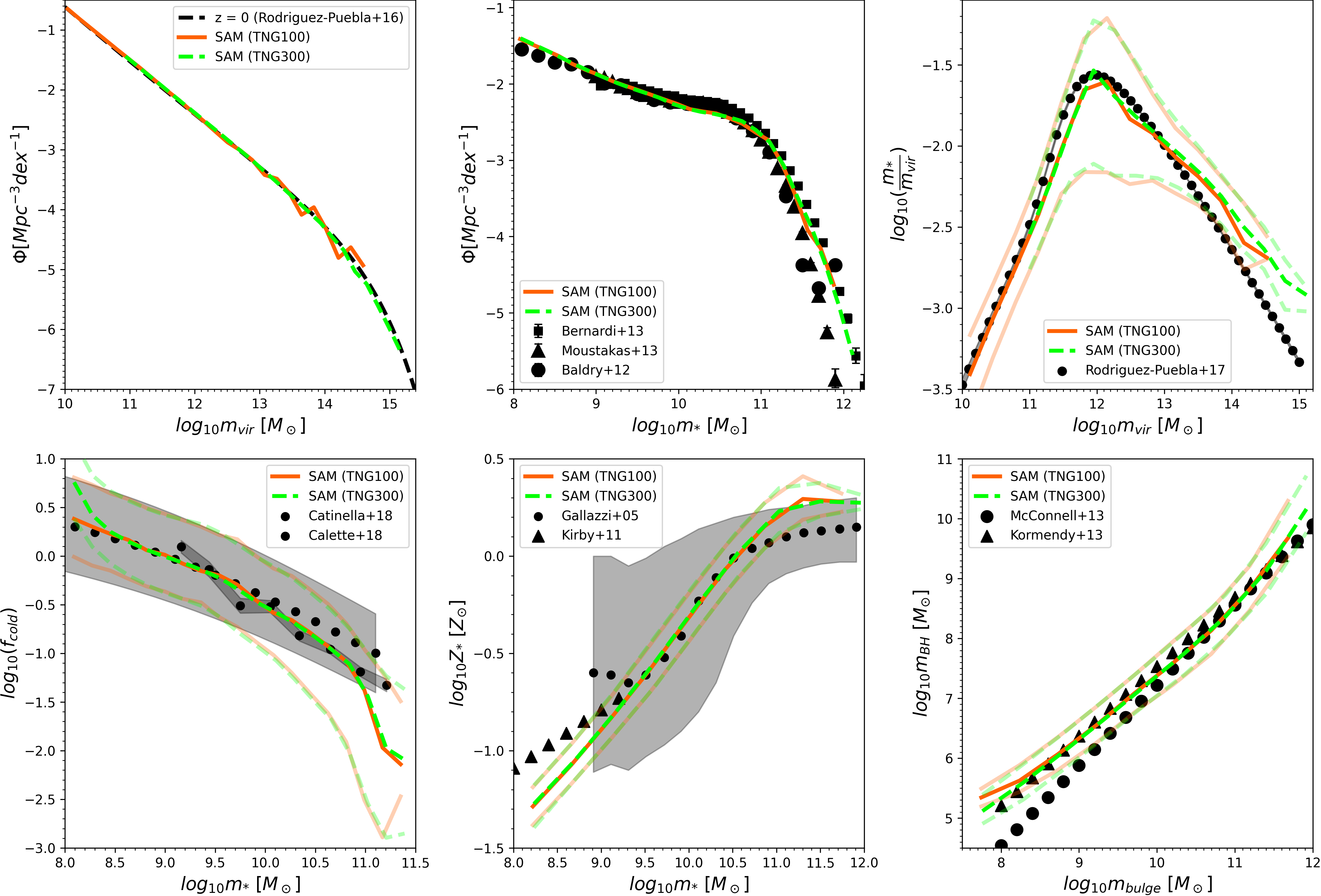}
    \caption{Suite of SAM calibration plots for 
    TNG100-1-Dark (orange) and TNG300-1-Dark (green) catalogs. (Top row) halo mass function (left), stellar mass function (middle), stellar mass-halo mass relationship (right). (Bottom row) cold gas fraction vs. stellar mass for disk-dominated galaxies (left), stellar metallicity vs. stellar mass (middle), black hole mass vs. bulge mass (right). In the top right and bottom panels, the opaque lines shows the medians and transparent lines show the 16 and 84th percentiles. The stellar mass function is for satellite and central galaxies, whereas all other relationships are for centrals only. A full complete list of references for the observational data used in the calibration is given in \citet{Yung:2019a}}. 
    \label{fig:calibration-suite}
\end{figure*}

We ran the SC SAM on the merger tree suites from 
TNG100 and TNG300 volumes. We require root halos to contain at least 1000 dark matter particles in order to ensure a robust merger history and measured properties -- if a root halo does not meet this threshold, it is excluded from further analysis. As noted above, the agreement between the results when the SC SAM is run on the TNG-based merger trees and those from the Bolshoi Planck merger trees is excellent, with no required recalibration. The adjustable parameters in the physics recipes described in \ref{sec:models:SAM} are chosen to match a standard set of physical properties derived from $z=0$ galaxy observations, as shown in Fig.~\ref{fig:calibration-suite}. These include the stellar mass function, the ratio of stellar mass to halo mass vs. halo mass (SMHM), the cold gas fraction (here defined as $f_{\rm cold} \equiv (m_{\rm HI} + m_{\rm H2}) / m_*$) vs. stellar mass for disk-dominated galaxies (defined as having stellar bulge to total mass ratio $B/T>0.4$), the stellar metallicity vs. stellar mass, and the black hole mass vs. bulge mass. Note that the halo mass function shown here is based on the \rockstar  halo catalog. This illustrates the range of halo masses that are represented in the TNG100 and TNG300 boxes. 

One can see that the agreement between the model predictions and the calibration quantities is in most cases very good. The stellar mass function has an offset of $\pm \sim 0.1$ dex with respect to the various scatters \citep{Baldry2012, Bernardi2013, Moustakas2013}. The SMHM scatter provided in \cite{Rodriguez-Puebla2017} fits within the 16th-84th percentile range of our calculated scaling relationship up until $\mvir \sim 10^{14} \msun$, where the tail ends of the 100 and 300 box relationships have offsets of $\sim 0.25$ and $\sim 0.4$ dex respectively. Similar behavior is seen in the $f_{cold}$ vs. $m_*$ scaling relationship - the scatter obtained from \cite{Catinella2018} and \cite{Calette2018} fit the calculated stellar mass medians for both boxes up until $m_* \sim 10^{11} \msun $, where there is at most an offset of $\sim 0.7$ dex towards the tail end of the relationships, still being within the calculated $84^{th}$ percentile. The stellar metallicity relationships, while not as good as the aforementioned relationships, still have good fits. The observational derived estimates provided by \cite{Gallazzi2005} and \cite{Kirby2011} fit well with the calculated medians between $10^{9} \msun \lesssim m_* \lesssim 10^{10.5} \msun$, whereas low and high stellar mass galaxies in both simulation boxes have offsets at most $\sim -0.25$ and $\sim +0.1$ dex respectively. The BH mass vs. bulge mass relation also has good agreement, although it is somewhat resolution dependent in the regime where the black holes are sensitive to the halo mass at which they are seeded.  It is also notable that the convergence is excellent -- with no recalibration of parameters, the predictions for the two TNG boxes, which vary in mass resolution by an order of magnitude, are nearly indistinguishable.
The predictions of the TNG simulations are not quite so well converged, with stellar masses at fixed halo mass in the TNG300 boxes predicted to be about 40\% lower across the board than in the TNG100 box (see Appendix A1, \citealp{pillepich:2018}).
In order to avoid the complication of attempting to place the predictions from the different TNG volumes on a self-consistent footing, in the present work we focus our comparison of the SAMs with just the TNG100 simulation. 

We note that TNG has similarly been calibrated to reproduce a set of observationally derived quantities, and although there is overlap between the calibration quantities used for the SAM and those used for TNG, there are some observations used by one technique that are not used by the other, and vice versa, and the observational studies adopted in the calibration also differ in some cases (see Section~\ref{sec:models:TNG}). We emphasize that the SAMs were calibrated according to the method used in previous published papers by the Santa Cruz group \citep{Somerville:1999, somerville:2001, somerville:2008, somerville:2012, PST:2014, somerville:2015}, and were not in any way retuned to match the results of TNG. 

\subsection{Statistical comparison of galaxy property distribution functions} 
\begin{figure*}
	\includegraphics[width=\textwidth]{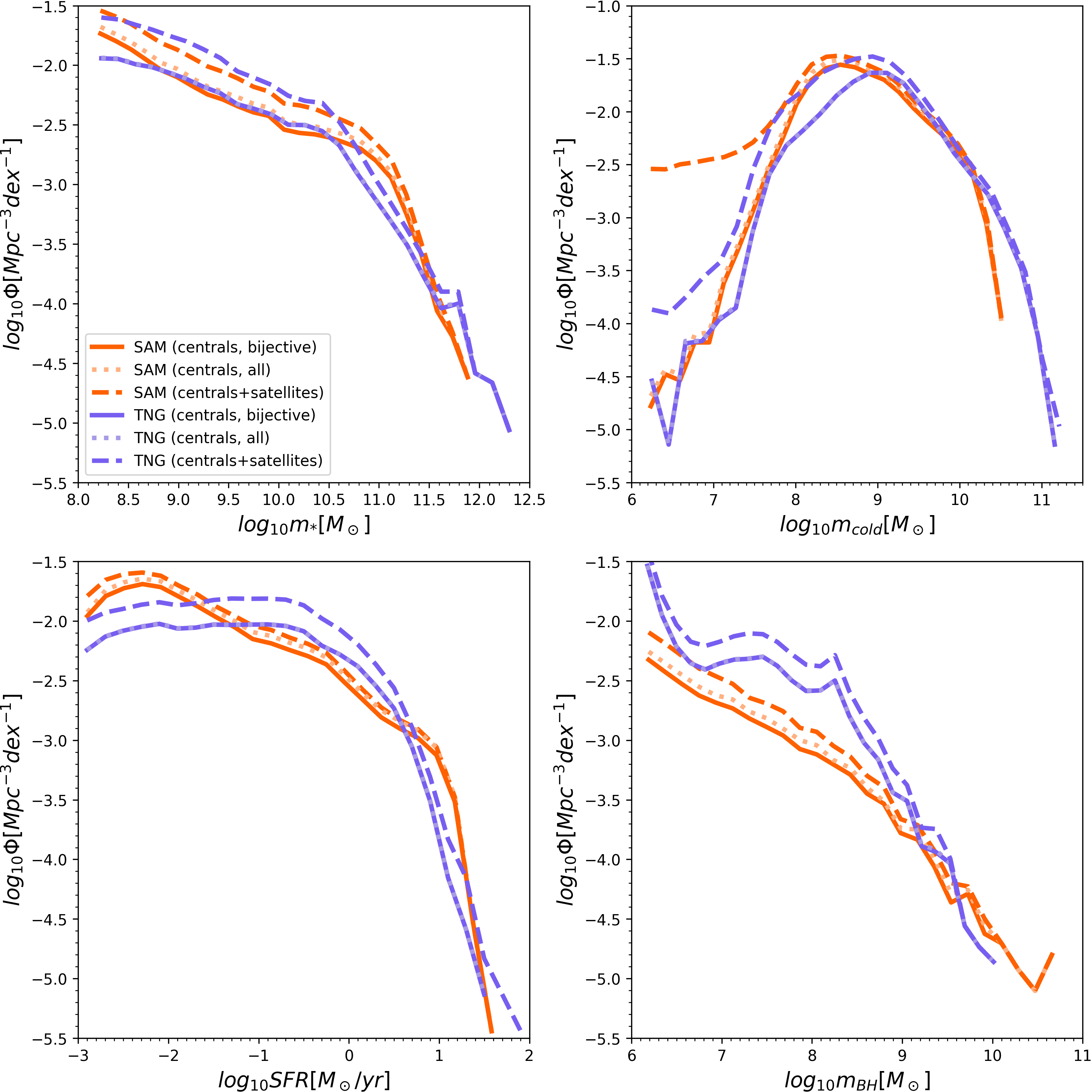}
    \caption{Comparison of $z=0$ functions for galaxy physical properties between the SC SAM (orange) and TNG (purple), where in both cases only galaxies in halos with virial halo mass of $M_{\rm vir} > 10^{10.5} \msun$ are included. From left to right: stellar mass function, black hole mass function, cold gas mass function, star formation rate function. The dashed lines represent the central + satellite  population, the dotted lines represent all centrals, and the solid lines represent bijectively matched centrals. Note that the resolution criteria described in \ref{tab:fields} have been applied to the population shown here.
    }
    \label{fig:mass-functions}
\end{figure*}

In Fig.~\ref{fig:mass-functions} we examine distribution functions (the number density of galaxies as a function of various physical properties) for the galaxies in both models. We show comparisons for three samples from the SAM and TNG: all galaxies (central and satellites), all central galaxies (regardless of whether they have a match), and central galaxies with bijective matches only. In all three cases, we are showing the sample population that passes the criteria described in Table \ref{tab:fields}. As we saw before, the majority of central galaxies have a bijective match, so that the bijective matched sample provides a fair representation of the overall population. There are instances though where a bijective subhalo may pass the resolution criteria for one model and not the other.
We compare the mass functions for the two models for a set of key observable quantities --- stellar mass, cold gas mass, SFR, and black hole mass. Our analysis specifically compares the centrals + satellites distribution functions for both models. Note that our sample is comprised of host halos where $m_{vir} > 10^{10.5} \msun$, where we define $m_{vir}$ for the SAM as $\mvir$ from the \rockstar catalog, and for TNG as m200c from the Subfind catalog.

The stellar mass functions for the two models agree very well within 0.1 dex, with some disagreement ($\sim 0.3$ dex) right around the knee ($ 10^{10.5} \msun /$ yr$ \lesssim m_* \lesssim 10^{11.5} \msun /$ yr). TNG predicts a slightly higher number density of massive galaxies with $m_{*} \gtrsim 10^{11.5} \msun$, the mass function reaching stellar masses greater than $10^{12}$ \msun. 
The cold gas mass functions (where $m_{\rm cold} = m_{\rm HI} + m_{\rm H2}$) agree well near the knee ($ 10^{9} \msun /$ yr $\lesssim m_{cold} \lesssim 10^{10} \msun /$ yr) with a difference of at most $\sim 0.1$ dex, but the SAM cold gas mass function drops off a bit faster at the high mass end. TNG has $\sim 10$ times as many galaxies than the SAM where $m_{cold} \sim 10^{10.5} \msun$ and has galaxies that have a cold gas mass content of $10^{11} \msun$. It should be kept in mind that the post-processed cold gas catalogs for TNG are incomplete at gas masses below $2 \times 10^{8} \msun$.
The SFR function shows reasonable agreement, although TNG has a higher normalization around the knee ($ 10^{-1} \msun / $ yr $\lesssim SFR \lesssim 1 \msun$ / yr) and the SAM predicts a higher number density of galaxies with low SFR. The ``knee'' in the SFR function in the SAM appears to occur at a higher SFR, around $1 \msun$ / yr.  
The black hole mass function shows the largest discrepancy, with TNG having a significantly higher number density of black holes with $m_{\rm BH} \lesssim 10^9 \msun$. We shall see later that this is a consequence of TNG growing black holes more efficiently in low mass halos. 

\subsection{Scaling relations, dispersions, and residuals} 

In this section, we investigate \emph{scaling relations} for our sample of bijective matched central galaxies. We define a scaling relation as the dependence of one galaxy property on another galaxy or halo property. 
We wish to address several questions: 
\begin{enumerate}
    \item How well do the \emph{median} scaling relations between halo and galaxy properties agree between the SC-SAM and TNG, in a statistical sense?
    \item How do the \emph{dispersions} in these scaling relations compare in a statistical sense?
    \item How do specific quantities compare halo by halo?
    \item Are the residuals from scaling relations correlated between the SC SAM and TNG on a halo by halo basis?
\end{enumerate}
 We perform this comparison for the same suite of galaxy properties described above ($m_*$, $m_{\rm cold}$, SFR, and $m_{\rm BH}$) as well as for hot circumgalactic medium gas, $m_{\rm hot}$. Note that we focus on scaling relations between these quantities and halo mass (as halo mass can be more easily linked between the two methods), but in Appendix~\ref{sec:suppscale} we also show variants of some of the results as a function of stellar mass, which may be more directly compared with observations. 

\begin{figure*}
	\includegraphics[width=\textwidth]{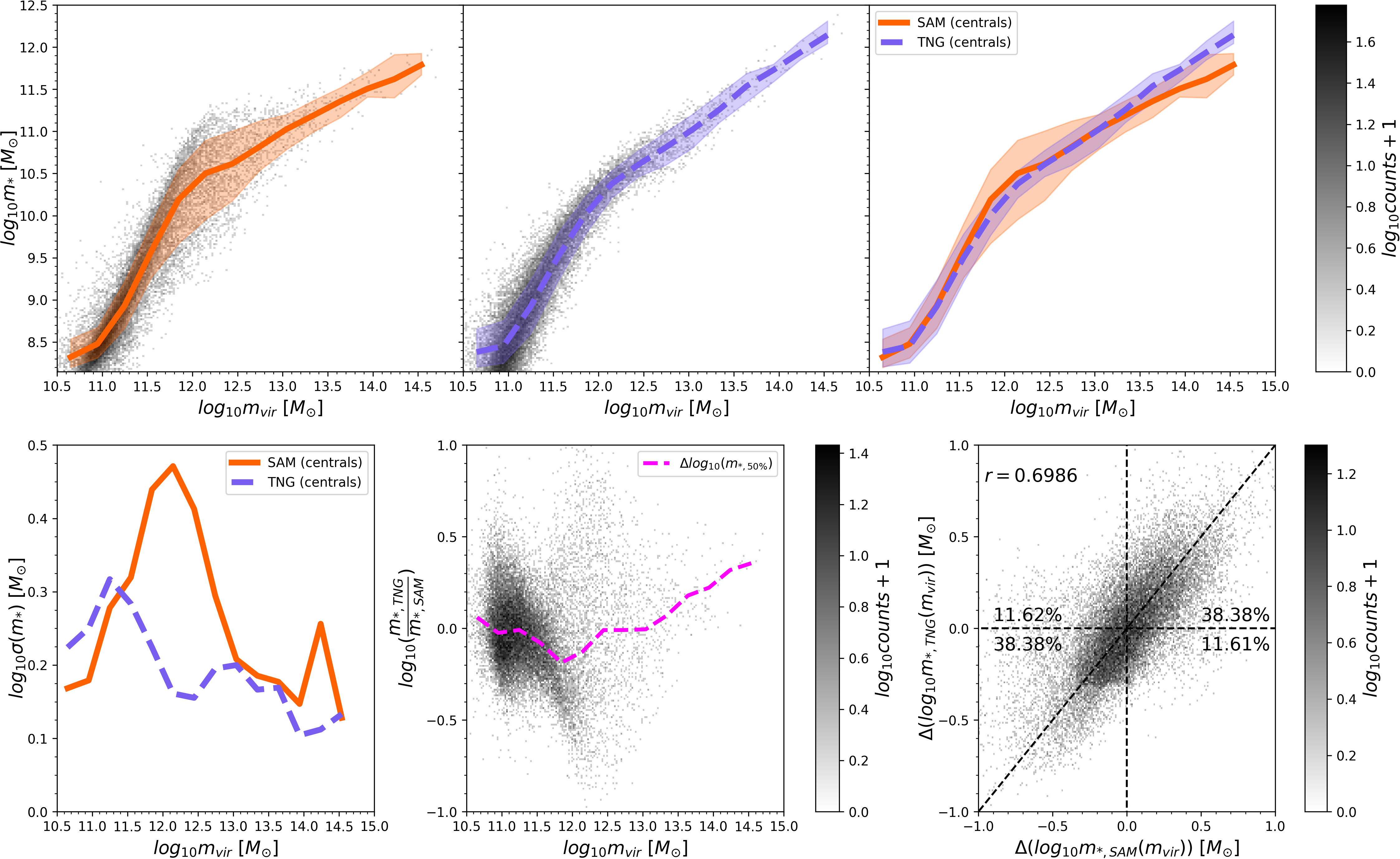}
    \caption{Stellar mass-halo mass (SMHM) scaling relation $m_*$ vs. $\mvir$. From left to right, top row: SAM density distribution (gray shading) with median (solid line) and 16 and 84th percentiles (orange shaded region), TNG density distribution with median and 16 and 84th percentiles (purple shaded region), SAM and TNG median and percentiles overplotted; bottom row: SAM and TNG dispersions, log of the ratio between $m_*$ in TNG and the SAM for matched halos, correlation between residuals from the median relation for TNG vs. the SAM for matched halos. In the bottom right panel, we indicate the value of the linear correlation coefficient $r$ and the fraction of objects in each of the four quadrants. }
    \label{fig:SMHM}
\end{figure*}

We first walk through the structure of a figure that will repeat for our selected series of quantities throughout this section. We can consider Fig.~\ref{fig:SMHM} as an example. On the top row we show the median and 16 and 84th percentiles for the SAM, TNG, and the two models overplotted. The values of all quantities that fall below the minimum value as determined by the resolution of the TNG simulation (see Appendix~\ref{sec:quantities} and Table~\ref{tab:fields}) are set to the minimum value, both for plotting purposes and for calculating the medians, percentiles, and residuals. 
On the bottom row, we show in the first panel the dispersion defined as 
$\sigma(q) = \frac{q_{84\%} - q_{16\%}}{2}$
as a function of halo mass 
for the SAM and TNG compared (if the distribution is Gaussian and symmetric, this would be equivalent to 1-$\sigma$). In the middle panel we show the halo-by-halo ratio of the property value as predicted by the SAM relative to that predicted by TNG vs. halo mass for the matched set of halos. In the right panel, we show the residual correlations, which are computed as follows: for bins in a parameter $x$, we compute the distance of the property value $y$ for a particular halo from the median value at that value of $x$. As a concrete example, let us take the stellar mass $m_*$. We compute $\Delta$ $m_{*, \rm SAM}(\mvir)$ by subtracting the value of $m_*$ for a given halo from the median value in the SAM at the relevant halo mass bin for $\mvir$. We repeat the same procedure for the TNG halo sample, now using the median from TNG to compute $\Delta$ $m_{*, \rm TNG}(\mvir)$. We then plot $\Delta$ $m_{*, \rm SAM}(\mvir)$ vs. $\Delta$ $m_{*, \rm TNG}(\mvir)$ 
for the full sample of halos. Objects in this plot can fall into one of four quadrants --– above the median in both models (upper right), below the median in both models (bottom left), above the median in the SAM and below the median in TNG (bottom right), and below the median in the SAM and above the median in TNG (upper left).  This diagram can provide valuable insights about whether halos that are above or below the median scaling relation in TNG are also above or below it in the SAM. 
Annotated on the plot are the percentage of objects that fall into each of the four quadrants and the value of Pearson's correlation coefficient, $r$. In this work, we consider the following ranges of $ r = |0, 0.2|, |0.2, 0.4|, |0.4, 0.6|, |0.6, 0.8|,$ and $|0.8, 1.0|$ to have no, weak, moderate, strong, and very strong correlations respectively. A correlation is considered `significant' if it at least falls into the moderate range. We remind the reader that the results shown throughout this section are only for central galaxies with bijective matches and that we define $\mvir$ to be the halo virial mass given by the Rockstar catalog in the dark matter only simulation.

We begin by investigating the scaling relationship between halo virial mass and stellar mass. Comparing the median relations, shown in the top row of Fig.~\ref{fig:SMHM}, we find excellent agreement within $\pm 0.1$ dex from the lowest mass halos that are well resolved ($\mvir \gtrsim  10^{11} \msun$) up to a halo mass of about $10^{12} \msun$, where the SAM predicts slightly higher stellar masses (by up to about 0.2 dex) over a fairly narrow halo mass range. 
Then, above $\mvir \sim 10^{13} \msun$, galaxies in TNG have higher stellar masses than those in the SAM, by an amount that increases with halo mass up to a maximum value of $\sim 0.5$ dex. 
Significant differences are seen in the dispersion in the SMHM relation (bottom row of Fig.~\ref{fig:SMHM}, leftmost panel). 
At the lowest halo masses, the dispersion in the TNG SMHM relation is higher
than that of the SAM, though it decreases rapidly, crossing the SAM value of the dispersion at $\sim 10^{11.5} \msun$ and continuing to decrease nearly monotonically up to the highest halo masses. Conversely, the dispersion in the SMHM relation of the SAM peaks strongly around  
$10^{12.25} \msun$, coinciding with the peak in the median scaling relation, and is quite similar to that in TNG for halo masses $\gtrsim 10^{13.25} \msun$. 

The halo-by-halo comparison (middle and right bottom panels in Fig.~\ref{fig:SMHM}) is consistent with the comparison of the statistical scaling relations seen in the top panels. Namely, the SAM predicts higher stellar masses in halos between $10^{11.5} \msun$ and $10^{12.25} \msun$. At halo masses above $10^{13.5} \msun$, TNG predicts higher galaxy stellar masses. The scatter in the halo-by-halo comparison is quite large --- up to 1.5 dex. A particularly interesting result is revealed by the residual correlation plot (rightmost bottom panel of Fig.~\ref{fig:SMHM}). A significant positive correlation is seen ($r =$ $0.70$) when comparing the SMHM residuals for TNG against those of the SAM. We label this correlation as strong based on our classification system. This indicates that the majority of halos that are above the median in the SAM are also above it in TNG, and vice versa, which hints at a similar relationship between halo formation history and stellar mass assembly in the two models. This is also demonstrated by examining the fraction of objects in each of the quadrants. There are significantly higher fractions in the ``correlated'' quadrants (38.39\% and 38.38\%) than in the ``uncorrelated'' quadrants (11.63\% and 11.62\%). 

\begin{figure*}
	\includegraphics[width=\textwidth]{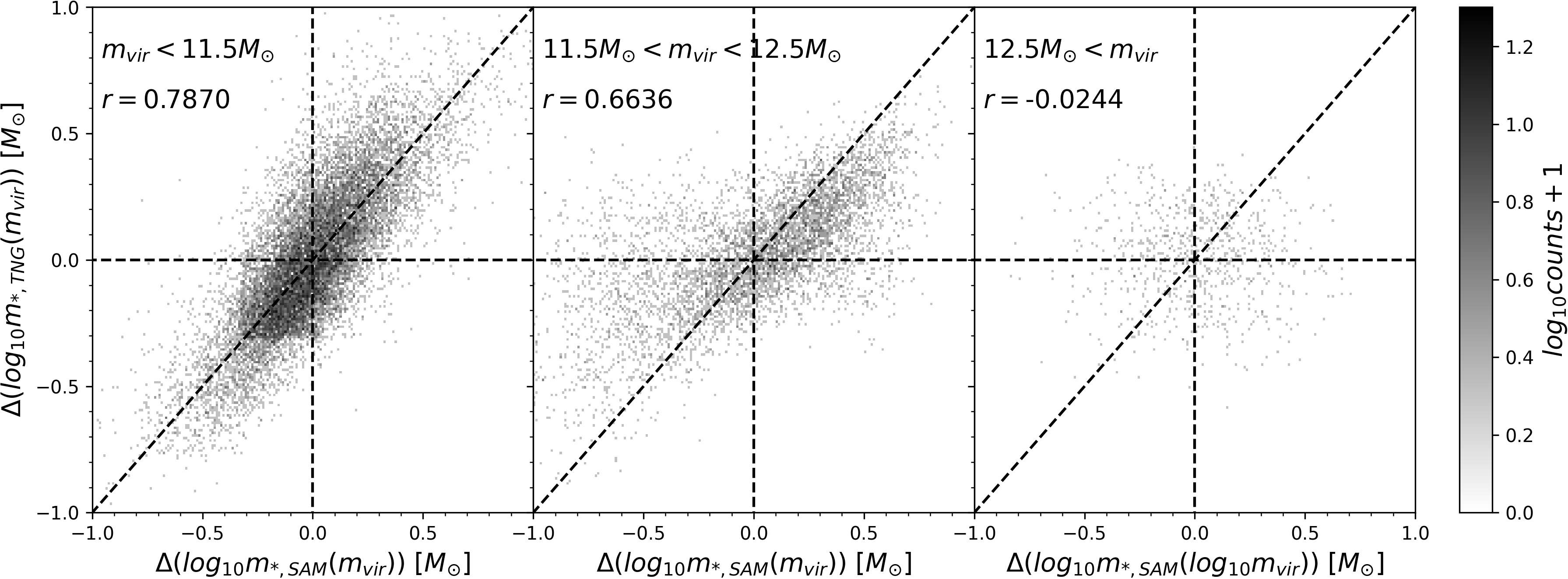}
    \caption{Residuals from the median SMHM relation in TNG and the SAM for matched halos, in different halo mass ranges. From left to right: $\mvir < 10^{11.5} \msun$, $10^{11.5} \msun < \mvir < 10^{12.5} \msun$, $10^{12.5} \msun < \mvir$. We see a stronger correlation in the lower halo mass bins.}
    \label{fig:SMHM-breakdown}
\end{figure*}

We investigate this further by plotting the same kind of halo-by-halo residual diagram for different mass bin ranges. 
In Fig.~\ref{fig:SMHM-breakdown}, we compare the SMHM residuals at low ($\mvir < 10^{11.5}$), median ($10^{11.5} \msun < \mvir <  10^{12.5} \msun$), and high ($10^{12.5} \msun <  \mvir$) virial masses. We obtain linear correlation coefficients of $r = $ $0.79$ (strong), $0.66$ (strong), $-0.02$ (no) for these mass ranges respectively. Thus the correlation breaks down for halo masses above the peak in the SMHM relation, which is also where star formation typically becomes quenched. We plan on further investigating this in a a follow up work where we track star formation and residual evolution across multiple redshifts (Gabrielpillai et al. in prep.).

\begin{figure*}
	\includegraphics[width=\textwidth]{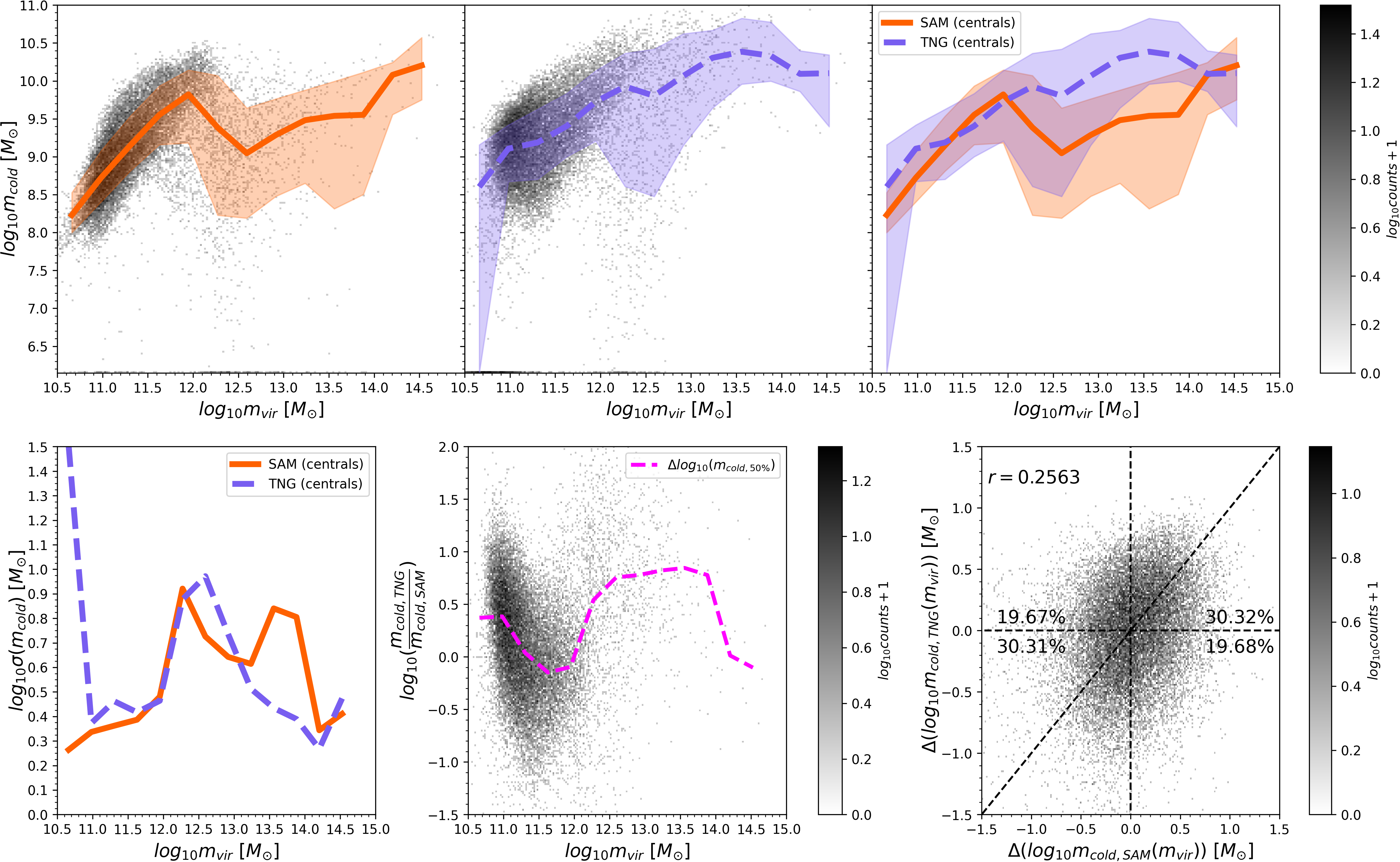}
    \caption{Scaling relation between cold gas mass $m_{\rm cold}$ and halo virial mass. Panels are as in Fig.~\ref{fig:SMHM}. }
    \label{fig:fcold-v-mhalo}
\end{figure*}

The next galaxy property we investigate is the cold gas mass $m_{\rm cold}$ as a function of $\mvir$, shown in Fig.~\ref{fig:fcold-v-mhalo}. Note that if a halo has a cold gas mass that is below the minimum value that can be resolved by TNG (as given in Table~\ref{tab:fields}), it is displayed along the bottom edge of the plot. In both the SAM and TNG, in any halos with $m_{\rm cold} < m_{\rm cold, min}$, $m_{\rm cold}$ is set equal to $m_{\rm cold, min}$ for purposes of calculating the dispersions and residuals. We define the cold gas mass as the mass of cold neutral gas in the central galaxy $m_{\rm cold} = m_{\rm HI} + m_{H2}$.

Examining the two scaling relations shown in the top row, we see that the SAM and TNG halos have similar cold gas masses below halo masses of $10^{12} \msun$, although TNG shows somewhat higher values at the lower end of this range. The SAM predicts a sharper drop in cold gas mass, starting at halo masses of around $10^{12} \msun$, while the predicted decrease is more gradual for TNG, such that the cold gas masses in TNG in the halo mass range $10^{12} - 10^{14} \msun$ are about a dex higher than in the SAM. However, the two models converge again to similar gas masses at the highest halo masses, $10^{14} - 10^{14.5} \msun$.

Interestingly, unlike for the previous SMHM relation, the dispersion in the cold gas content of halos is almost identical in the SAM and TNG in the low- and medium-mass range. Above a halo mass of $10^{13.5} \msun$, the SAM shows a sharp increase in the dispersion, but this is driven by a small number of halos and may reflect the onset of a bimodality in the distribution as galaxies become depleted of cold gas by AGN feedback. However, the SAM does seem to predict a larger number of massive halos with extremely low cold gas content than does TNG.

The halo by halo comparison shown in the bottom middle panel of Fig.~\ref{fig:fcold-v-mhalo} shows that the predicted cold gas mass frequently differs by up to $\pm 1$-1.5 dex, and shows the trend that high mass halos ($\mvir \gtrsim 10^{12} \msun$) in TNG have median values of cold gas mass about an order of magnitude higher than the corresponding halos in the SAM. 

In contrast with the SMHM relation, a less significant correlation is found between the halo by halo residuals from the median $m_{\rm cold}-\mvir$ relation (correlation coefficient of $r = $ $0.26$) with a very similar number of objects in each quadrant. We examined this relationship at different mass ranges as in the SMHM case, but found no significant correlation in any of the mass bins. A version of Fig.~\ref{fig:fcold-v-mhalo} shown as a function of stellar mass instead of halo mass is provided in Appendix~\ref{sec:suppscale} (Fig.~\ref{fig:fcold-v-mstar}) where we also find a weak correlation in the residuals. 

\begin{figure*}
	\includegraphics[width=\textwidth]{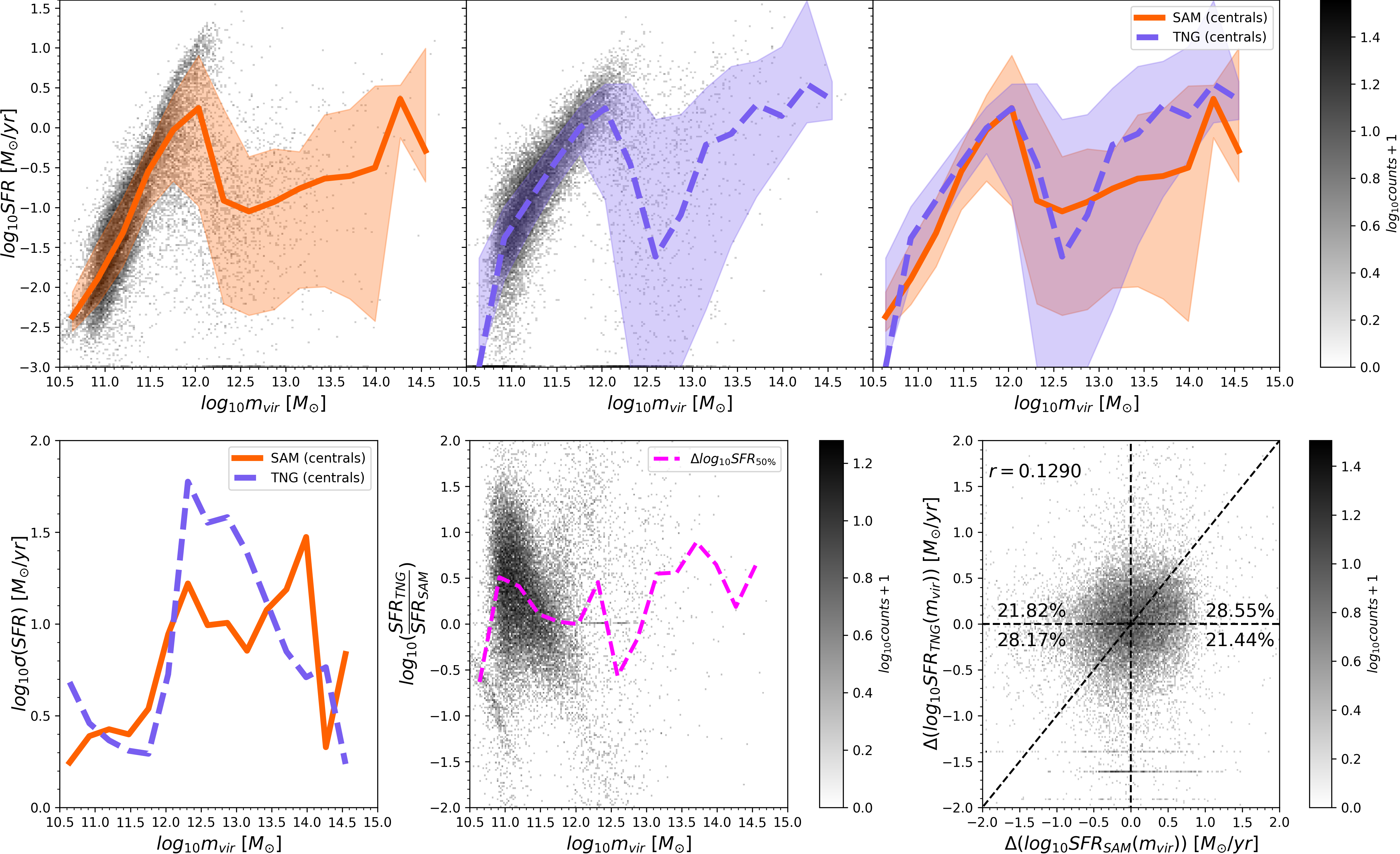}
    \caption{Scaling relations between star formation rate SFR vs. halo mass. Panels are as in Fig.~\ref{fig:SMHM}.  }
    \label{fig:sSFR-v-mhalo}
\end{figure*}

A galaxy's star formation rate is closely tied to its cold gas reservoir, so we next consider a similar analysis of the star formation rate (SFR) as a function of $\mvir$, shown in Fig.~\ref{fig:sSFR-v-mhalo}. As in the previous figure, galaxies with SFR values smaller than minimum value that can be resolved in TNG have their SFR set equal to the minimum SFR, and are displayed at the bottom edge of the plot.
Beginning with the median scaling relationship, we can see that the two models predict qualitatively similar behavior, with a nearly linearly increasing SFR with stellar mass below a halo mass of $\sim 10^{12} \msun$, and a sharp decline above this mass, that is commonly termed quenching. Despite this similarity, though, TNG galaxies consistently have a higher star formation rate than SAM galaxies both below and above the quenching threshold, with the exception around $10^{12.5} \msun$.  This is consistent with the higher cold gas fractions that were also seen in TNG in Fig.~\ref{fig:fcold-v-mhalo}.

Moving to the dispersion, the SAM and TNG show fairly similar dispersions below the quenching mass of $\mvir \sim 10^{12} \msun$. TNG predicts a higher dispersion in the halo mass range $10^{12.25} - 10^{13.5} \msun$. Both models show a similar dispersion in SFR for the highest halo masses $\gtrsim 10^{14} \msun$. As expected from the scaling relations, the halo by halo comparison shows that the SFR in the SAM galaxies are systematically about 0.2--0.5 dex lower than those in TNG, with an overall scatter of around $\pm 1$ dex. As for $m_{\rm cold}$, there is no significant correlation found between the halo by halo residuals from the median SFR vs. halo mass relation ($r = $ $0.13$). In fact, the correlation is even weaker than the one seen in $m_{\rm cold}$, which is not surprising given that the instantaneous SFR is expected to be somewhat stochastic. A version of Fig.~\ref{fig:sSFR-v-mhalo} shown as a function of stellar mass instead of halo mass is provided in Appendix~\ref{sec:suppscale} (Fig.~\ref{fig:sSFR-v-mstar}) where we see a weak correlation in the residuals.

\begin{figure*}
	\includegraphics[width=\textwidth]{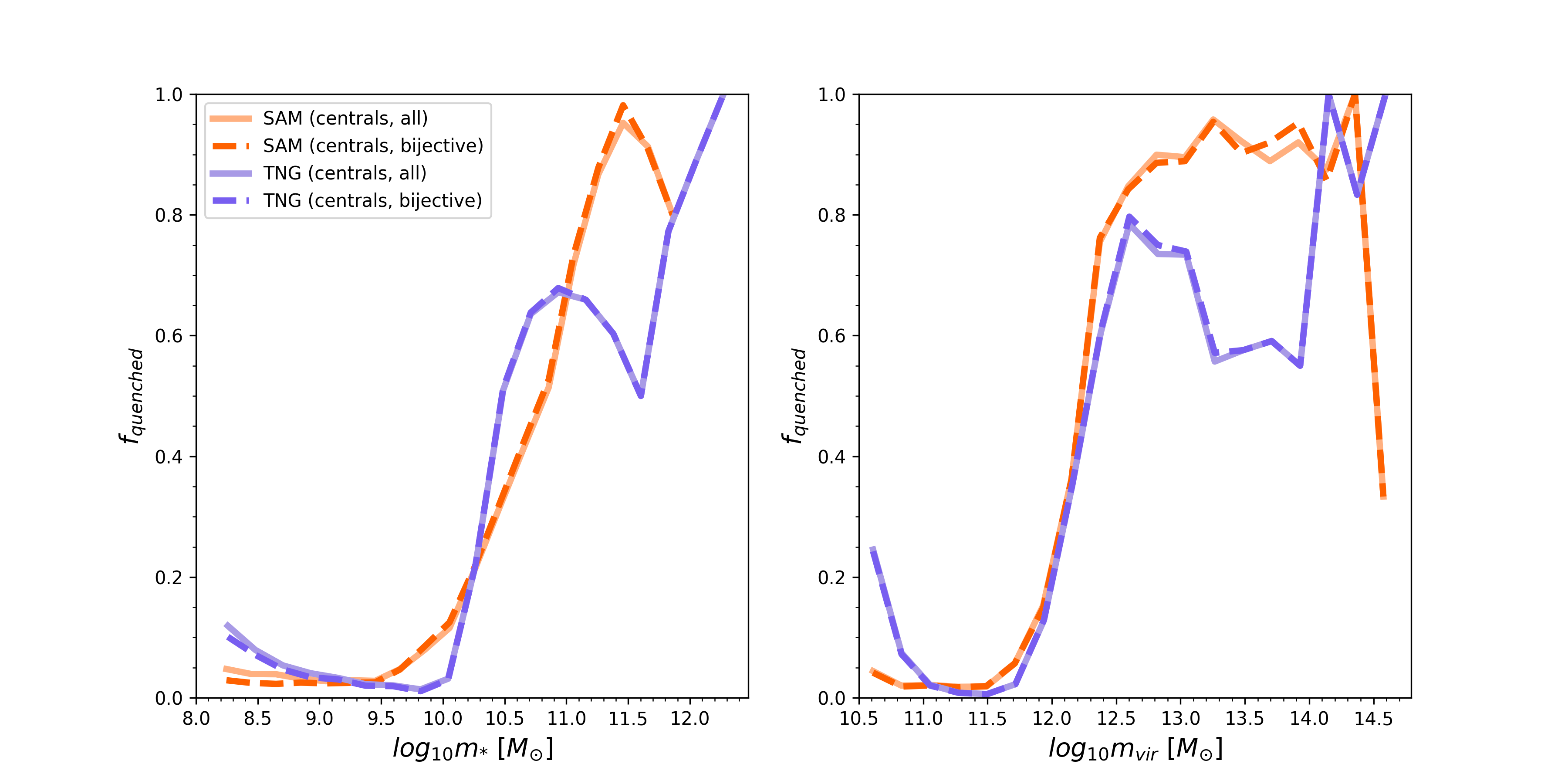}
    \caption{(left) Fraction of quenched objects as a function of stellar mass. (right) Fraction of quenched objects as a function of halo mass. The solid line depicts all central subhalos whereas the dashed line represents bijectively matched centrals. The predicted quenched fractions agree quite well between the SC SAM and TNG for low- and median-halo masses and stellar masses}.
    \label{fig:nquenched}
\end{figure*}

We further investigate star formation activity and quenching in the two models by comparing the fraction of quenched galaxies as a function of both stellar and halo mass (Fig.~\ref{fig:nquenched}).  For this work, we consider a galaxy to be star forming when its specific star formation rate (sSFR), defined as sSFR = SFR / m$_*$, satisfies the condition sSFR$ > 10^{-11} $yr$^{-1}$ and to be quenched when  sSFR$ < 10^{-11} $yr$^{-1}$.  Overall, the qualitative agreement between the SAM and TNG in the quenched fraction is quite good Although there is a 20\% discrepancy between the SAM and TNG where $\mvir < 10^{11} \msun$, the two models have a very similar quenched fraction nearly on top of each other until beginning to diverge at $\mvir \sim 10^{12.3} \msun$. The two still follow a consistent trend, diverging at most by 10\% until $\mvir \sim 10^{12.3} \msun$ where TNG experiences a sharp drop in its fraction, agreeing again for high mass halos where $\mvir \sim 10^{14} \msun$. $f_{quenched}$ as a function of stellar mass shows a similar story - agreement within 15\% where $m_{*} \sim  10^{11} \msun$ where there is a  drop in the TNG fraction until the two models converge again where $m_{*} \sim  10^{11.8} \msun$ Both models show very low quenched fractions below a transition mass of $m_* \sim 10^{10} \msun$ or $\mvir \sim 10^{11.7} \msun$, with TNG however predicting somewhat higher quenched fractions in the very lowest mass galaxies/halos than the SAM. These may be splashback halos, halos that became stripped of their gas by passing through a larger halo, but are currently considered central galaxies. The SAM cannot accurately track splashback halos, so such halos would be considered satellite galaxies. Alternatively, these galaxies could be affected by other environmental effects that are not included in the SAM.

Both models show a rapid increase in quenched fraction at around the same stellar mass $(m_* \sim 10^{10} \msun$) or halo mass $(\mvir \sim 10^{12} \msun$), and plateau at a quenched fraction of around 90\%. TNG shows a dip in the quenched fraction at around galaxy stellar masses of $10^{11.5} \msun$ or halo masses of $10^{13}-10^{14} \msun$. This feature has been seen in other studies \citep[e.g.][]{ayromlou:2021}, and likely is related to the critical BH mass that triggers the onset of the kinetic mode of AGN feedback \citep[see][for details]{weinberger:2017}. Comparing these results with those seen in Fig.~\ref{fig:sSFR-v-mhalo}, it is interesting that although there is little correlation between whether an individual halo is quenched or star forming in the SAM and TNG, nonetheless, the statistical properties of the population are quite similar. 

\begin{figure*}
	\includegraphics[width=\textwidth]{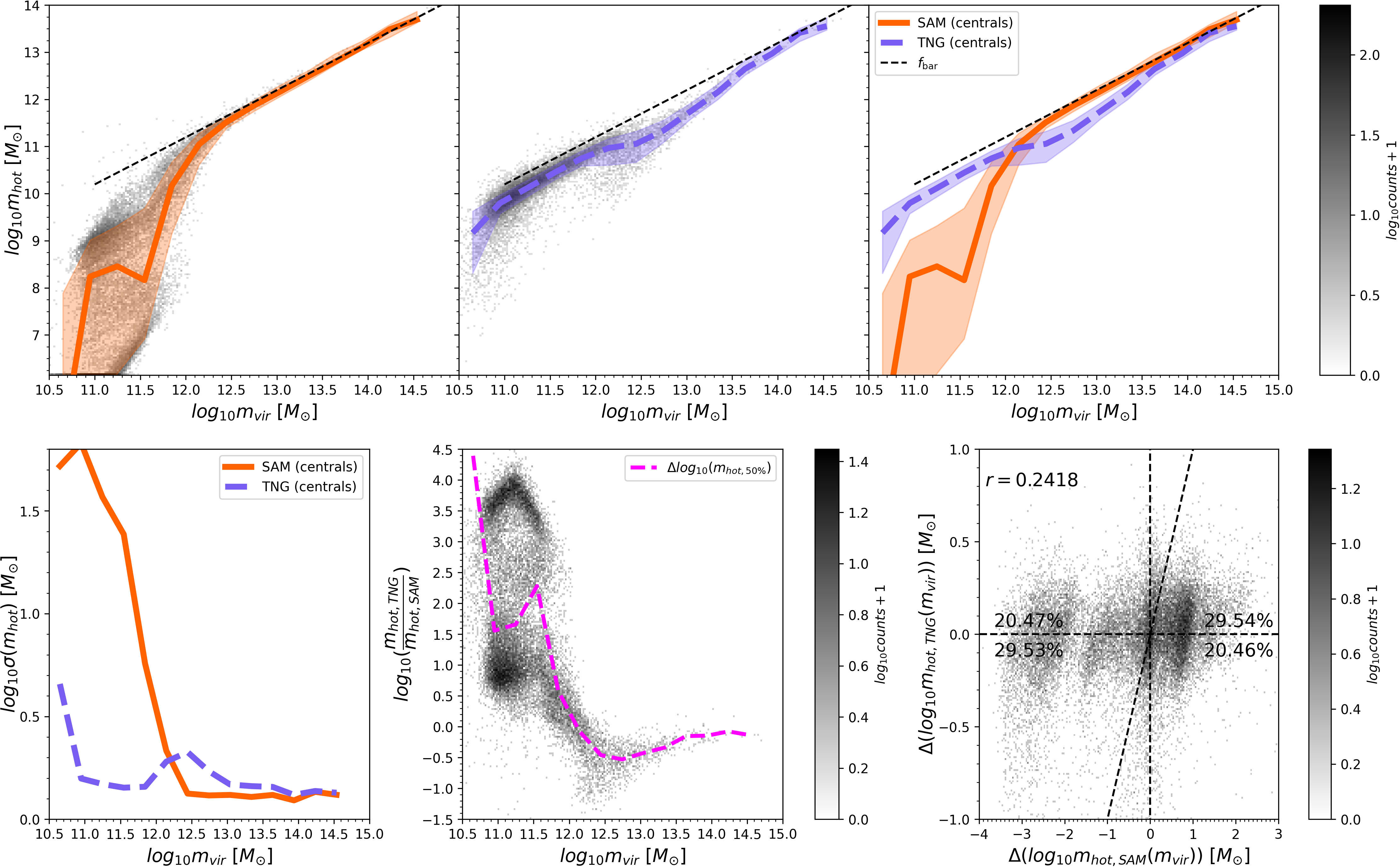}
    \caption{Scaling relations between hot gas (CGM) mass $m_{\rm hot}$ vs. halo mass. Panels are as in Fig.~\ref{fig:SMHM}.  }
    \label{fig:fhot-fraction}
\end{figure*}

We continue our investigation in Fig.~\ref{fig:fhot-fraction} by analyzing the circumgalactic medium (CGM) and by comparing the hot gas mass $m_{\rm hot}$. 
We compute $m_{\rm hot}$ in TNG as 
\begin{equation}
    \label{eq:mhot}
 m_{\rm hot} = m_{\rm group} - \sum_{i=1}^{N_{\rm sub}} m_{\rm cold, i}
\end{equation}
where $m_{\rm group}$ is the total gas mass in the FoF group (GroupMassType (Type=0)) and $m_{\rm cold, i}$ are the cold gas masses of each subhalo within that FoF group\footnote{Note that the gas mass field in TNG includes wind particles.}.

As in previous figures, where the gas mass in a halo in either the TNG or the SAM is less than the minimum mass specified in Table~\ref{tab:fields}, we set the value equal to the minimum mass both for display purposes and for computing the dispersions and residuals. 
The median relations for $m_{\rm hot}$ vs. $\mvir$ probably show the worst agreement of any of the quantities that we have considered, which is perhaps not surprising, considering the dearth of observational constraints on this quantity for halo masses lower than group scales. The predicted hot gas masses in the SAM are orders of magnitude lower than TNG for halo masses below $10^{12} \msun$. Moreover, in this mass range, the SAM predicts a bimodal or multi-modal distribution of halo gas content, which is not seen in TNG. The agreement at higher halo masses is quite good being within 0.6 dex, especially if one disregards the slight dip at around $\mvir \sim 10^{12}$--$10^{13.5} \msun$ in TNG, which is again caused by the BH mass dependent transition to the kinetic AGN feedback mode. 

Examining the dispersion, shown in the bottom left panel of Fig.~\ref{fig:fhot-fraction}, we see that both the SAM and TNG have a very small dispersion in hot gas mass for halo masses above $10^{12} \msun$, with TNG showing a slight increase in dispersion around the same halo mass range where the dip in the median relation appears. Strikingly, the dispersion in hot gas mass remains quite low for TNG all the way to the lowest halo masses, while it blows up dramatically in the SAM, reaching over 1.5 orders of magnitude higher values at the lowest halo masses we consider. From the middle panel, showing the halo by halo comparison of $m_{\rm hot}$, we again see that the SAM is producing a bimodal/multimodal population of halos, some with very low hot gas mass, which is part of the reason for the extreme discrepancy seen in the median relation and the dispersion. However, even the higher gas mass population at low halo masses has $m_{\rm hot}$ values that are about 1 dex lower than those in TNG. Finally, a weak correlation is seen between the residuals from the median $m_{\rm hot}$-$\mvir$ relation in the matched halo population in the SAM and TNG  ($r = 0.18$). We observe that there is a moderate correlation between the residuals for high virial mass objects and a weak one for lower mass objects.

The possible physical reasons for these discrepancies are discussed in Section~\ref{sec:discussion}. 

\begin{figure*}
	\includegraphics[width=\textwidth]{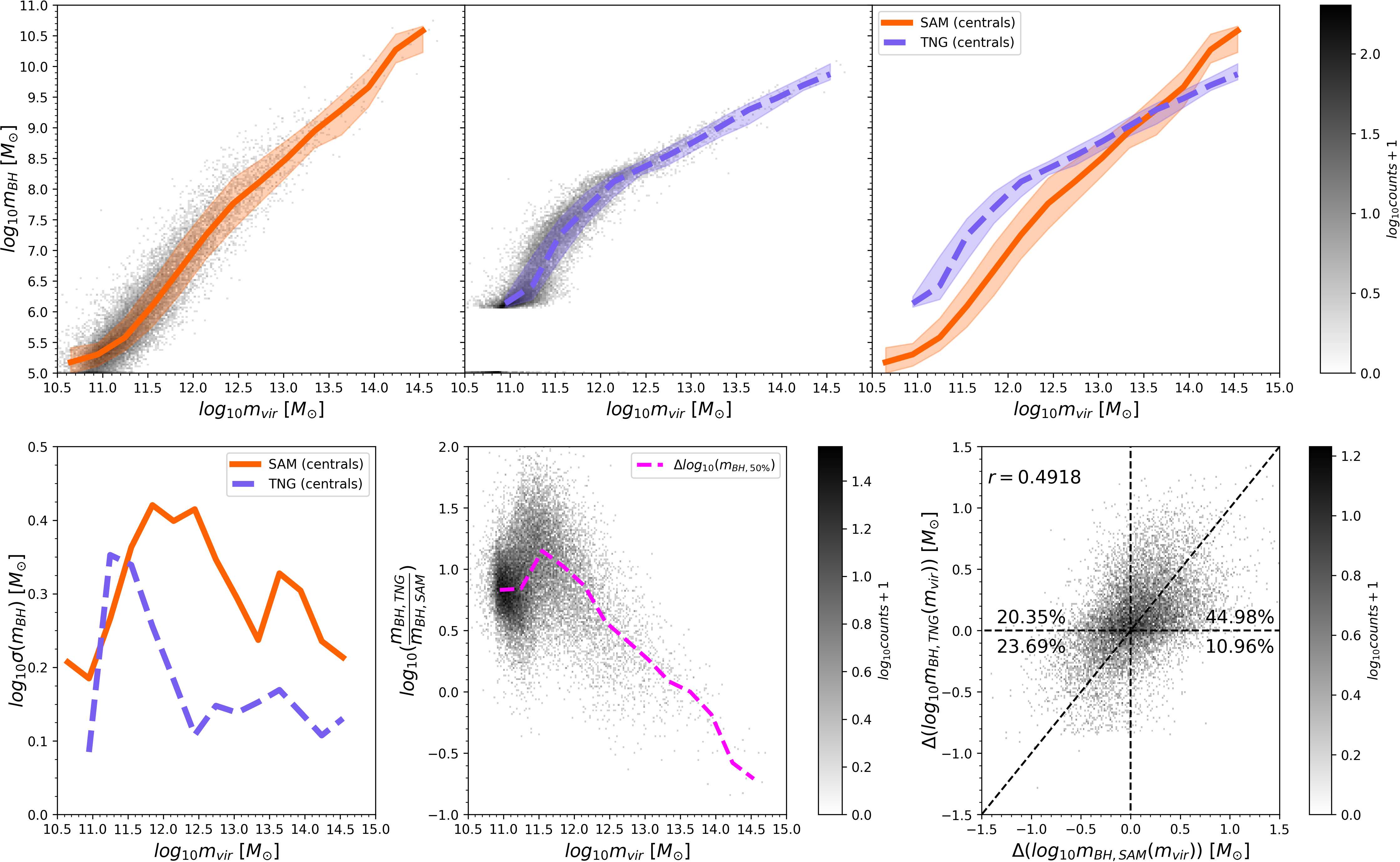}
    \caption{Scaling relations between BH mass $m_{\rm BH}$ vs. halo mass. Panels are as in Fig.~\ref{fig:SMHM}. }
    \label{fig:BH-fraction}
\end{figure*}

The final scaling relation we consider is between BH mass and the mass of the host halo, shown in Fig.~\ref{fig:BH-fraction}. Here the SAM and TNG show very different results for the median relations. The BH seed mass in TNG is $1.25 \times 10^6\, \msun$, while it is $1.0 \times 10^4\, \msun$ in the SAM. The relationship between BH mass and halo mass is also much steeper in the SAM, so that high mass halos $\mvir \gtrsim 10^{13.5}\, \msun$ have more massive BH in the SAM than TNG and the reverse is true for halos with masses less than this. The shape of the $m_{\rm BH}$ vs. $\mvir$ relation is very similar to that of the SMHM relation, especially in TNG, indicating that there is a nearly linear relationship between stellar mass and BH mass (see Fig.\ref{fig:mBH-v-mstar}). 

The dispersion in $m_{\rm BH}$ at fixed halo mass also shows significant differences in the two models. Note that in halos with masses below $\sim 10^{11} \msun$, the different values of the seed BH mass dominate the results, so the decrease in dispersion is not very physically significant.  The dispersion in $m_{\rm BH}$ in the SAM increases rather continuously with decreasing halo mass across the full mass range, until it reaches the halo mass scale where the seed mass begins to affect the results. TNG, in constrast, has a lower dispersion of $\sim 0.15$ dex from $\mvir \sim 10^{12.5}$--$10^{14.5}\, \msun$, and a strong increase in dispersion to lower halo masses below $\mvir \sim 10^{12.5} \msun$. 

The halo by halo comparison of the bijective matched sample shows a similar trend as what could be inferred from the comparison of the median relations; namely, that the BH mass is about an order of magnitude higher in TNG at the lowest halos masses we analyzed, and about 0.5 dex lower than the SAM at the highest halo masses. The residual-residual plot shows a moderate correlation between the SAM and TNG ($r = 0.49$). We find that the correlation remains moderate when separating the population into low, medium, and high virial mass bins as previously mentioned. A version of Fig.~\ref{fig:BH-fraction} shown as a function of stellar mass instead of halo mass is provided in Appendix~\ref{sec:suppscale} (Fig.~\ref{fig:mBH-v-mstar}), where no significant correlation is observed in the residuals.

\section{Discussion}
\label{sec:discussion}

\subsection{Comparison with other work}

There is an abundant literature on comparing the predictions of semi-analytic models with those of hydrodynamic simulations. For example, \citet{SD15} show a statistical comparison of several key galaxy property distributions, and their redshift evolution, including three hydrodynamic simulations and five SAMs. They found that these models showed close agreement for quantities that are commonly used in the calibration procedure (such as the stellar mass function at $z=0$), but diverged more for higher redshifts and for quantities such as cold gas fraction and gas phase metallicity. Their comparison showed that the differences between one hydrodynamic simulation and another, or one SAM and another, are as large as those between hydrodynamic simulations and SAMs -- namely, there was no clear systematic difference between the predictions of numerical hydrodynamic simulations and SAMs in the quantities that they investigated. 

In the following discussion, we focus on several recent comparisons that have involved either the Santa Cruz SAM or the TNG simulations, or both. Perhaps the most relevant to our study is the work of \citet{ayromlou:2021}. Similar to our study, they performed a statistical and a halo-by-halo comparison of the predictions of the {\sc l-galaxies} SAM \citep{Henriques:2015} with TNG. The {\sc l-galaxies} SAM contains many similar ingredients to the SC SAM, and the techniques have many aspects in common, but the details of the physics ingredients as well as the calibration are different. \citet{ayromlou:2021} show several of the same quantities that we examine here, so a direct comparison is possible. They find that the median predicted stellar-to-halo mass relation at $z=0$ agrees between {\sc l-galaxies} and TNG at the level of better than 40 percent, with a 1-$\sigma$ dispersion in the halo-to-halo difference between TNG and {\sc l-galaxies} of around 0.2-0.3 dex. These results are qualitatively quite similar to our findings. They also find that the discrepancies between the SAM and TNG are the largest around the peak in the SMHM relation ($\mvir \sim 10^{12} \msun$), and that TNG predicts larger SMHM ratios than {\sc l-galaxies} above this halo mass. These results are both similar to what we find as well. 
Another quantity that can be compared directly is the SFR or specific star formation rate (see Fig.~\ref{fig:sSFR-v-mstar}). At low galaxy masses, \citet{ayromlou:2021} find that TNG predicts slightly lower sSFR than {\sc l-galaxies}. This is the reverse of what we find, in which the SC SAM predicts lower SFR and sSFR than TNG for low-mass galaxies.
The sSFR in TNG show a very strong dip relative to {\sc l-galaxies} at stellar masses of $\sim 10^{10.8} \msun$. We see a hint of a similar feature in the SC SAM vs. TNG comparison, but it is not as pronounced.

Also directly comparable are the quenched fractions. Here, the SC SAM shows close agreement with TNG in the critical mass at which the quenched fraction drops, while the quenched fraction for central galaxies in {\sc l-galaxies} drops more gradually and starting from a higher stellar mass than TNG or the SC SAM. Finally, we can also compare the cold and hot gas content. \citet{ayromlou:2021} find that TNG galaxies have less cold gas than {\sc l-galaxies} galaxies across all masses, by around 0.3 dex at low masses, increasing to 2.5 dex at around the quenching mass, and by about 1.5 dex at high masses ($\sim 10^{11.5} \msun$). We find overall somewhat better agreement between cold gas content in the SC SAM and TNG overall, and find the opposite trend at high masses (TNG has higher cold gas masses than the SC SAM, but lower than {\sc l-galaxies}). The comparison of hot gas mass shows qualitatively very similar trends: \citet{ayromlou:2021} find much lower hot gas masses in low mass halos ($\mvir \lesssim 10^{12} \msun$) in {\sc l-galaxies} relative to TNG, similar to although not to quite as great an extent as what we find in the SC SAM. They also see a dip in the hot gas mass in TNG between $\mvir \sim 10^{12}$--$10^{14} \msun$, which is not seen in either SAM, and in both SAMs the hot gas masses at $\mvir \gtrsim 10^{12} \msun$ are larger than those in TNG. 

Another relevant recent study is that of \citet{Pandya:2020}, which performed halo-by-halo comparisons between the SC SAM and the FIRE-2 hydrodynamic simulations. They found fairly good agreement between the predictions of these two models for the stellar mass and cold gas mass fractions vs. halo mass, but found that the SC SAM underpredicts the hot gas (CGM) masses in FIRE-2 by several orders of magnitude. Moreover, \citet{Pandya:2020} found that the physical processes regulating the formation of stars in the SC SAM and FIRE-2 are fundamentally different --- in FIRE-2, there is significant \emph{preventative feedback} in low mass halos that prevents gas from cooling and collapsing into the ISM. In the SC SAM, gas inflow rates in low mass halos are much larger than those in FIRE-2, requiring much larger mass \emph{outflow} rates in order to end up with similar stellar fractions. 

One of the exciting possible applications of SAMs is to predict galaxy clustering properties in large volumes that would be too expensive for numerical hydrodynamic simulations. There have been a few recent studies that appear encouraging in this regard. \citet{renneby:2020} compared predictions of galaxy-galaxy lensing profiles and clustering for the {\sc l-galaxies} SAM and TNG with observations. They found that TNG produced better agreement with these observables, but they were able to modify the {\sc l-galaxies} SAM to improve the agreement with observations and with TNG. \citet{boryana:2021} compared predictions of galaxy clustering and assembly bias in the SC SAM and TNG, finding encouraging levels of agreement between both statistics. Furthermore, they studied the correlation of secondary halo properties, such as halo concentration and environment, on assembly bias. They find that the SC SAM shows a qualitatively similar response of galaxy occupancy and clustering to these secondary parameters as TNG.

\subsection{Interpretation of the results}

It is well known that many of the galaxy properties predicted by semi-analytic models and hydrodynamic simulations agree with one another to about the same extent as those predicted by different hydrodynamic simulations agree with one another \citep[e.g.][]{SD15}. The question at the heart of this work is: where the predictions of these two different techniques agree, is it due to common calibration practices, lucky coincidence, or is it an indication that similar physical processes are shaping galaxy properties in both models? Conversely, where they disagree, is it an indication of fundamentally different underlying physics or due to different calibration procedures? We may also define ``agreement'' or ``disagreement'' in different terms: in this work, we consider comparisons of statistical properties for the global galaxy population in both simulations and more rigorous halo-by-halo comparisons for bijective matched samples. We also consider not only agreement in median relations, but also dispersions and residual-residual correlations in scaling relations, which have not been studied in detail before.  

Schematically, the physical processes in the SAM and TNG have many common elements. For example, the mass loading of stellar driven winds is assumed to scale with an internal velocity on relatively small scales (in the SC SAM, the circular velocity at 2$r_s$ is used, where $r_s$ is the NFW scale radius; in TNG, the velocity dispersion in a weighted kernel over the nearest 64 DM particles is adopted). Star formation follows a Kennicutt-Schmidt type relation between cold gas density and star formation rate density. 
Supermassive black holes self-regulate in such a way that a relationship between BH mass and galaxy mass is enforced, and energetic feedback from jets driven by low-level accretion onto these black holes is able to heat hot circumgalactic gas and prevent it from cooling and providing new fuel for star formation in the central galaxy. That said, there are numerous differences in the details of how these physical processes are implemented in the two models, as well as in how the free parameters they contain are calibrated.  

The physical process that is most directly responsible for shaping the SMHM relation at low halos masses ($\lesssim 10^{12} \msun$) in both the SC SAM and TNG is stellar feedback \citep{somerville:2008,pillepich_TNGmethods:2018}.  The relatively good agreement between the SMHM relation in the SC SAM and TNG in this halo mass range suggests that this process may be operating in a similar manner in the two models.  

\begin{figure*}
	\includegraphics[width=\textwidth]{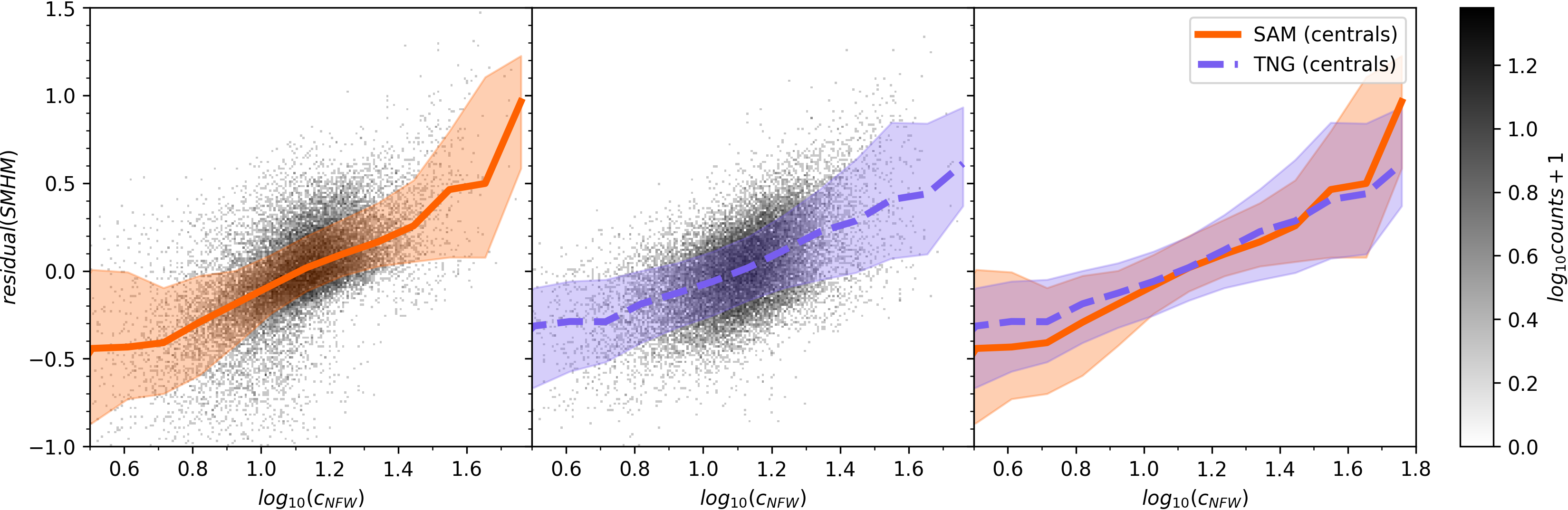}
    \caption{The difference between the $m_*/\mvir$ value for halos and the median value of $m_*/\mvir$ in a corresponding halo mass bin (SMHM residual) plotted as a function of the NFW concentration parameter of the halo. The left panel shows the relationship for the SC SAM, the middle panel shows that for TNG, and the rightmost panels shows a comparison. The solid lines show the medians and the shaded areas show the 16 and 84th percentiles. Both models show a correlation between $c_{\rm NFW}$ and SMHM residual, with a very similar slope. }
    \label{fig:cnfw}
\end{figure*}

However, it is also possible that the good agreement is simply a result of both models being calibrated to the same observational data at $z=0$, and could be achieved through a variety of different physical pathways, as seen for example in the study of \citet{Pandya:2020}. But the correlation between the residuals from the median relation for individual halos in TNG and the SAM more convincingly suggests that the connection between halo formation history and galaxy properties is similar in both models. To explore this further, we investigate the correlation between the residual from the SMHM median relation and the halo concentration parameter, which is known to be strongly correlated with halo formation history. This is shown for both models in Fig.~\ref{fig:cnfw}. A strong correlation is seen in both models, with a similar slope in the region where most halos are found. This same qualitative correlation between SMHM residual and halo concentration has been demonstrated in the EAGLE hydrodynamic simulations \citep{matthee:2017}. This may also help explain why the assembly bias signal is so similar in the SC SAM and TNG, as halo concentration (or assembly history) is known to be one of the most important secondary halo parameters that influences clustering \citep{boryana:2021}. In order to gain further insights into this question, in a follow-up paper we plan on investigating the stellar mass assembly histories as a function of time in TNG and the SC SAM (Gabrielpillai et al. in prep.). 

The underlying model for the star formation efficiency would also seem to be similar, as both models implement a variant of a Kennicutt-Schmidt relation, in which SFR density is a function of the gas surface density. However, the SC SAM implements a recipe for gas partitioning into molecular, atomic, and ionized gas, and star formation is a function of only the molecular gas component. Although there is a volume density threshold for star formation in TNG, this is not exactly equivalent to the gas partitioning recipe. Also, the galaxy size affects the surface density distribution in each object, and it is unknown whether the model implemented in the SC SAM (in which galaxy size depends on the halo specific angular momentum) predicts good agreement with the sizes in TNG. 

It is clear that the star formation rate and cold gas fraction depend more strongly on the recent conditions in the galaxy, while the stellar mass depends on the whole assembly history. Finally, it may be that the star formation rate and cold gas fraction depend largely on the recent infall rate of fresh gas, which depends on the conditions in the CGM, discussed below. 

The mismatch between the predicted CGM properties in the SC SAM and TNG in low mass halos is by far the most dramatic of any that we studied. The cooling model in the SAM is based on the classic picture presented by \citet{white-frenk:1991}, in which gas in low mass halos has a cooling time much shorter than the dynamical time, so that gas is assumed to fall in on a free-fall time. This is coupled with an assumption in the SC SAM that gas that is ejected from the ISM by stellar feedback is expelled all the way out of the hot halo and deposited in an ``ejected gas'' reservoir. As a result, many low mass halos in the SAM have essentially no hot gas in their CGM at all. This is in clear contradiction with the predictions of both TNG and FIRE-2. This strongly suggests that the cooling model and the fate of ejected gas from the ISM in the SC SAM should be revised to achieve better consistency with numerical simulations. 

The last major ingredient is the growth of supermassive black holes and quenching by AGN feedback. We found that SMBH grow much more efficiently in low mass halos in TNG than they do in the SC SAM. This may be because in TNG, there is no highly effective mechanisms for regulating black hole growth on small scales, while in the SC SAM, black hole growth is tied to the growth of the stellar bulge rather than the total galaxy stellar mass. Low mass galaxies are less efficient at forming bulges, so they are also less efficient at growing black holes in the SC SAM. 

Finally, although there is no significant correlation between the halo-by-halo SFR residuals in TNG and the SAM (i.e., a specific halo that is quenched in one model is not necessarily quenched in the other model), the fraction of quenched galaxies agrees statistically extremely well between the two models, 
with the onset of quenching occurring at the same stellar mass and halo mass. 

\section{Conclusions}
\label{sec:conclusions}

We have extracted halos (using the \rockstar\ halo finder) and halo merger trees (using the \ct\ merger tree builder) from the dark matter only runs of the 
TNG100-1-Dark and TNG300-1-Dark simulations, and run the published Santa Cruz semi-analytic models within these merger trees. The publicly available TNG catalogs make use of a different halo finder ({\sc subfind}), so we have built a sample of bijective matches between the \rockstar\ and \subfind\ halos for both the TNG dark matter only and full physics catalogs. We carried out statistical and halo-by-halo comparisons of a set of key galaxy physical processes in the two models that are important for understanding how galaxies form and assemble their stars and SMBH, and how star formation within them is regulated. These include the stellar mass function, cold gas mass function, and black hole mass function, and scaling relations between halo mass and stellar mass, cold gas mass, SFR, hot gas (CGM) mass, and BH mass. A novel feature of our study is that we have gone beyond comparison of the median scaling relations to study the dispersion and residual-residual correlations in the two models, which may yield greater insight into the underlying physical processes.  

Our key results include the following: 
\begin{itemize} 
\item The predicted distribution functions for stellar mass, cold gas mass, and SFR, are overall in good agreement for TNG and the SAM. The largest discrepancy is for the BH mass function, where the SAM predicts a significantly lower number density of low mass BHs than TNG.

\item The residual from the SMHM relation for bijective matched halos in TNG vs. that in the SAM shows a significant correlation. This correlation is strongest for low and intermediate mass halos, but disappears for massive halos ($\mvir \gtrsim 10^{12.5}\, \msun$). This implies that halos that have higher (lower) than average stellar mass in TNG tend to also have higher (lower) than average stellar mass in the SAM. 

\item The SMHM residual shows a strong correlation with halo concentration in both the TNG and SAM. The slope and normalization of this relation are very similar in both models. 

\item The median scaling relations for the cold gas mass and SFR show good qualitative agreement at low and medium virial halo masses, although the median cold gas content and SFR values in massive galaxies in TNG are offset to higher values relative to the SAM.

\item The dispersion in $m_{\rm cold}$ at fixed halo mass is similar in the SAM and TNG. The dispersion in SFR in TNG is much larger in the halo mass range just above the quenching mass than it is in the SAM.

\item The residual-residual relation for $m_{\rm cold}$ and SFR for matched halos shows no significant correlation. This implies that although the predicted medians and dispersions for these relations are very similar in the SAM and TNG for low and medium viral halo mass objects, an individual halo that has a higher (lower) than average value of $m_{\rm cold}$ or SFR in TNG does \emph{not} necessarily have a higher (lower) than average value of these parameters in the SAM.

\item The fraction of quenched central galaxies in the SAM and TNG as a function of stellar mass and halo mass shows very good qualitative agreement.

\item The hot gas (CGM) mass ($m_{\rm hot}$) shows the greatest discrepancy between the two models of all the quantities that we investigated, particularly at halo masses $\mvir \lesssim 10^{12} \msun$, where the medians disagree by as much as two orders of magnitude.

\item The SC SAM predicts a much steeper median BH mass vs. halo mass relation than TNG, implying that TNG is much less efficient at growing BH in low mass halos. 

{\item 
The dispersion in the BH mass vs. halo mass relation is much larger in the SAM than it is in TNG.}
 
\end{itemize}

In conclusion, the Santa Cruz SAM makes very similar predictions to TNG for statistical distributions and median scaling relations for several key galaxy properties, including stellar mass, cold gas mass, and SFR. The SAM also makes similar predictions for galaxy clustering and assembly bias as TNG \citep{boryana:2021}. These results are encouraging for the prospects of using SAMs as a tool to ``emulate'' the results of numerical hydrodynamic simulations, including for making predictions for galaxy clustering over large volumes. However, some properties (perhaps not co-incidentally, those that are less well constrained by observations), such as the CGM mass and the BH mass at low halo masses, show much larger mismatches in the two models. Moreover, for some properties (e.g. $m_{\rm cold}$ and SFR), even when the median relations and distribution functions appear to be in fairly good agreement, the halo-by-halo comparison reveals that the physical processes acting in both models must have very different dependence on halo formation histories, which calls into question whether quantities like the clustering properties for galaxies selected by this property will be in agreement. This has implications for using these methods to make forecasts for clustering of emission line selected galaxies or intensity mapping surveys that trace cold gas. Our study highlights physical processes (such as stellar feedback) that appear to have similar dependence on halo formation history in the SAM and TNG, and others (such as star formation and BH growth efficiency) that are very different. This will guide our future attempts to develop SAMs that can faithfully emulate the predictions of numerical hydrodynamic simulations for a broad suite of galaxy and halo properties.

\section*{Acknowledgements}
The material is based upon work supported by NASA under award number 80GSFC21M0002.
rss is supported by the Simons Foundation. We gratefully acknowledge the use of Flatiron Institute computing facilities for this work. 
The following Python software packages were utilized in this work: Python 3.7, NumPy, pandas \citep{mckinney-proc-scipy-2010, reback2020pandas}, h5py \citep{h5py}, matplotlib, and yt \citep{yt}.

\section*{Data Availability}

All data products used in this work will be made available through the IllustrisTNG website \url{https://www.tng-project.org/data/}.



\bibliographystyle{mnras}
\bibliography{paper} 




\appendix

\section{Supplementary scaling relationship plots}
\label{sec:suppscale}

In this appendix, we show figures that are similar to Fig.~\ref{fig:SMHM}-\ref{fig:sSFR-v-mhalo} and Fig.~\ref{fig:BH-fraction}, but we now plot $m_*$ on the x-axis. These quantities are more easily compared to the study of \citet{ayromlou:2021}, and are also more directly comparable with observations. The conclusions regarding the comparison of the median scaling relations, dispersions, and residuals drawn from these diagrams are qualitatively similar to those drawn from Fig.~\ref{fig:SMHM}-\ref{fig:sSFR-v-mhalo} and Fig.~\ref{fig:BH-fraction}.

\begin{figure*}
	\includegraphics[width=\textwidth]{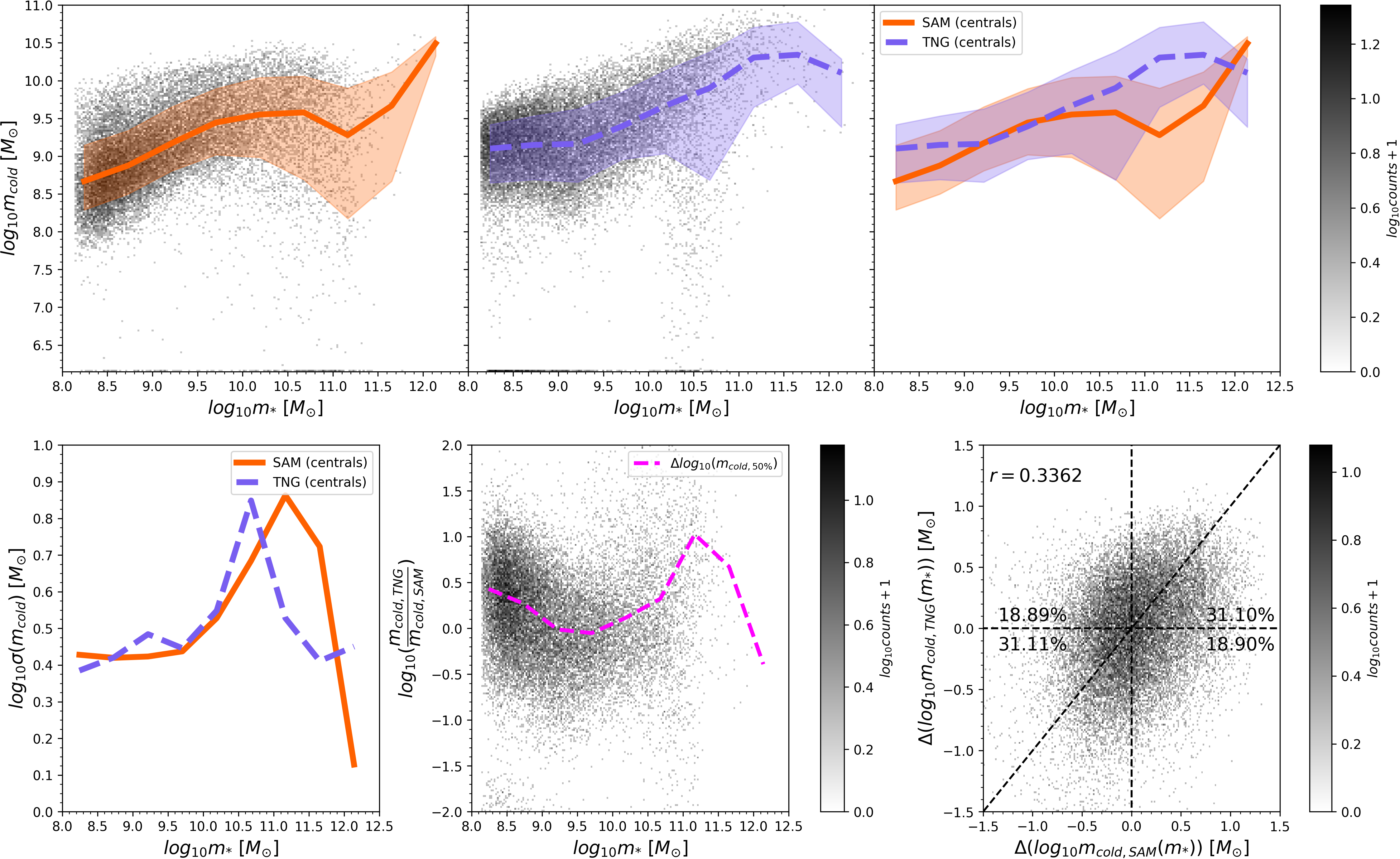}
    \caption{Scaling relations between cold gas mass and stellar mass for TNG and the SC SAM. Panels are as in Fig.~\ref{fig:SMHM}. }
    \label{fig:fcold-v-mstar}
\end{figure*}

\begin{figure*}
	\includegraphics[width=\textwidth]{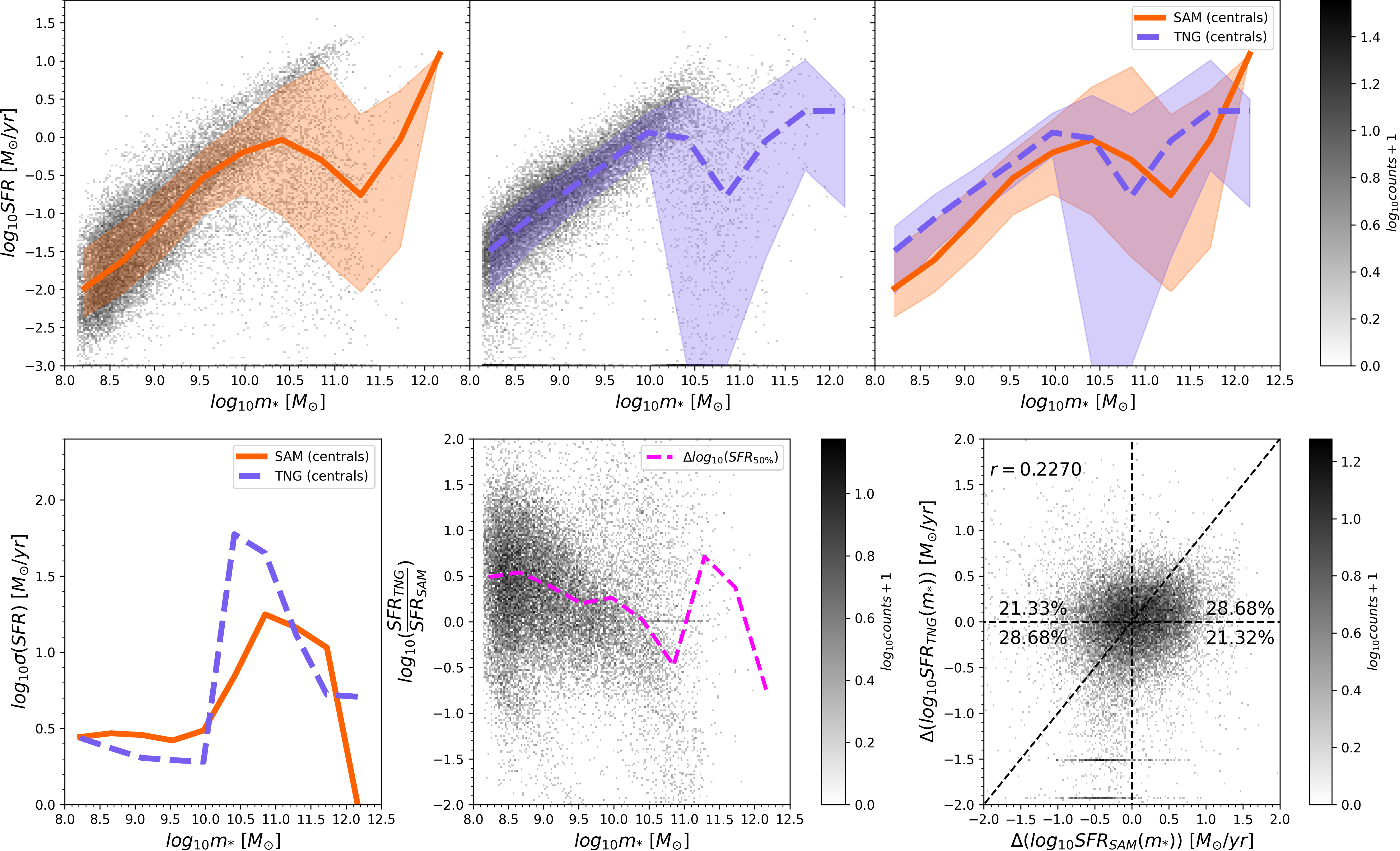}
    \caption{Scaling relations between SFR and stellar mass for TNG and the SC SAM. Panels are as in Fig.~\ref{fig:SMHM}.}
    \label{fig:sSFR-v-mstar}
\end{figure*}

\begin{figure*}
	\includegraphics[width=\textwidth]{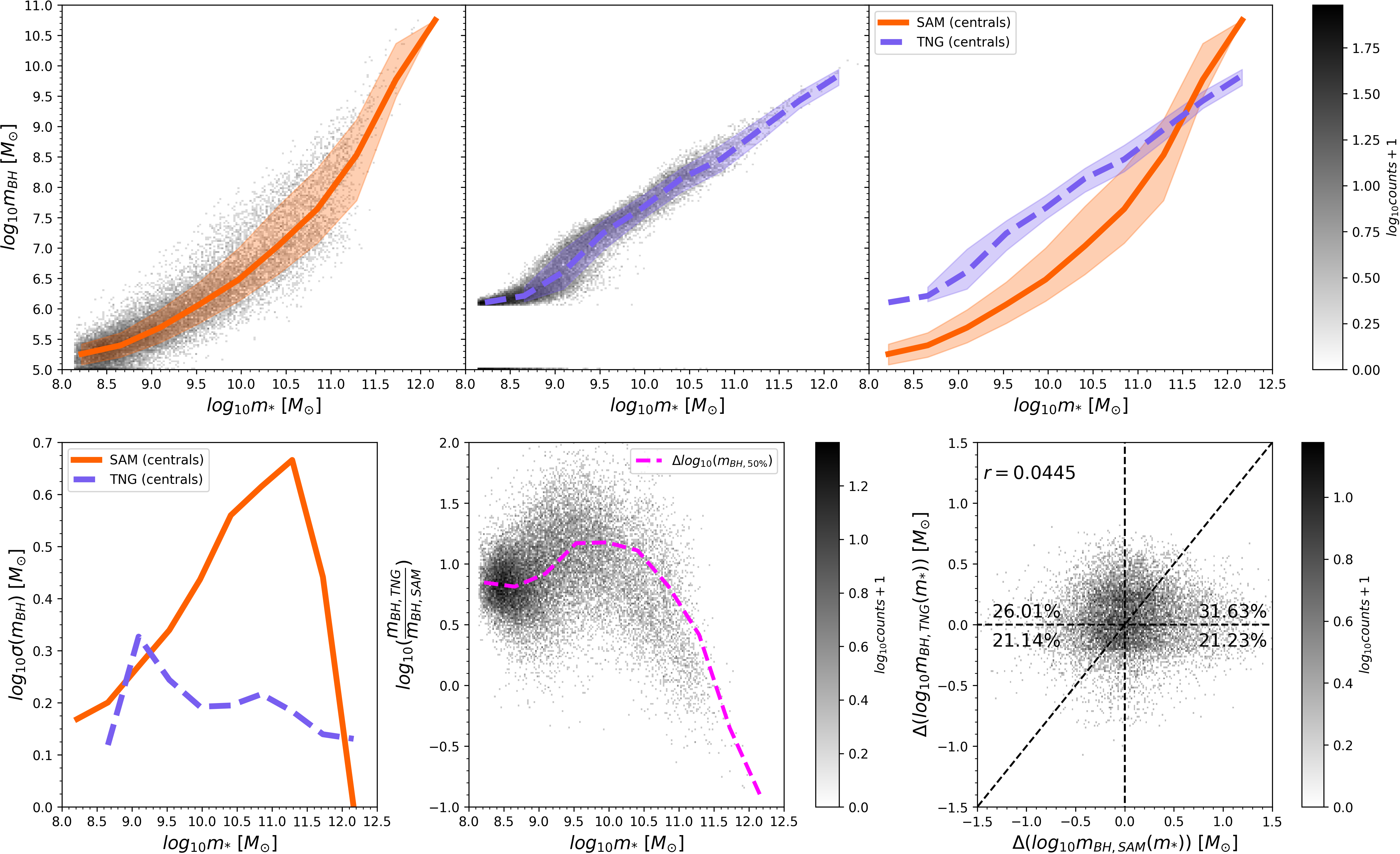}
    \caption{Scaling relations between BH mass and stellar mass for TNG and the SC SAM. Panels are as in Fig.~\ref{fig:SMHM}.}
    \label{fig:mBH-v-mstar}
\end{figure*}

\section{Guide to quantities used in our analysis}
\label{sec:quantities}

\begin{table*}
\centering
\caption{Field labels for TNG and SAM quantities used in this work.}
\label{tab:fields}
\begin{tabular}{llll}
\hline
Quantity & TNG label & SAM label & minimum value\\
\hline
$m_{\rm vir}$  & Group\_M\_TopHat200 & HalopropMvir & $8.9 \times 10^{10}\, \msun$\\
$m_{*}$  &  SubhaloMassInRadType (Type=4) & GalpropMstar & $1.4 \times 10^8\, \msun$ \\
$m_{\rm cold}$ & m\_hi\_GK11\_vol + m\_h2\_GK11\_vol & GalpropMHI + GalpropMH2 & $1.4 \times 10^6\, \msun$ \\
$m_{\rm hot}$  & see Eqn 2 & HalopropMhot & $1.4 \times 10^6\, \msun$  \\
SFR & SubhaloSFR &  GalpropSFR & $1.0 \times 10^{-3}\, \msun {\rm yr}^{-1}$  \\
$m_{\rm BH}$ & SubhaloMBH & GalpropMBH & $2 \times m_{\rm BH, seed} $ \\
		\hline
	\end{tabular}
\end{table*}

 Table~\ref{tab:fields} specifies the field labels for the galaxy and halo properties used in this analysis, as they appear in the TNG database. We have made an effort to compare the quantities that are the most analogous in the SAM and TNG, but we note that in some cases the quantities are not defined in exactly the same way. For example, the halo virial mass $m_{\rm vir}$ for TNG is the total mass contained within a sphere with a specified overdensity in the Full Physics TNG simulation, which contains baryons, while the SAM $m_{\rm vir}$ is defined in the same way, but is measured in the DM only simulation (see Fig.~\ref{fig:mvir-diff}). The stellar mass, $m_*$, is defined as the mass of star particles within twice the half mass radius in TNG, while in the SAM it is the total stellar mass within the galaxy. For very massive galaxies, there can be a significant amount of stellar mass outside of twice the the half-mass radius \citep{pillepich:2019}. However, in the SAM, $m_*$ does not include the mass of stars that may be scattered out of the main body of galaxies in mergers, nor the mass from tidally disrupted satellites, which may be major contributors to the stellar mass in extended envelopes in TNG. The fraction of cold ISM in the form of neutral hydrogen is computed in post-processing in TNG, while in the SAM it is computed and used in estimating the star formation rate self-consistently within the simulation. Moreover, the quoted cold gas mass is the total amount of gas bound to the subhalo, differing from the stellar mass estimate which is measured within twice the stellar half mass radius. In order to better match the gas mass estimate, we also use the SFR estimate within the whole subhalo for TNG. Both the TNG and SAM SFR estimates are instantaneous rates.  

Another fundamental difference between the predictions from TNG and the SAM is the finite mass resolution of TNG. The mass resolution of the SAM is set by the resolution of the DM simulation used to extract the merger trees, in this case $8.9 \times 10^6 \msun$ for TNG100-Dark. Using a rule of thumb that a halo should have at least 100 particles to be well resolved, and in order to resolve the merger history the root halo should be at least 100 times the mass of the lowest mass progenitor that can be resolved, a conservative minimum root halo mass is $\sim 8.9 \times 10^{10} \msun$. For TNG, the target baryon mass is $1.4 \times 10^6 \msun$, meaning that the mass of gas in each cell is kept within a factor of two of this mass, as is the mass of star particles at birth \citep{pillepich_TNGmethods:2018}. We restrict our analysis to TNG galaxies with a minimum of $\sim 100$ stellar particles ($m_{\rm star} > 1.4 \times 10^8 \msun$) where stellar mass is involved, and to galaxies/halos with at least one gas particle ($>1.4 \times 10^6 \msun$) where gas quantities are involved. Similarly, the lower limit on the gas mass per cell results in a lower limit on the SFR. We adopt a minimum SFR of $1.0 \times 10^{-3}\, \msun {\rm yr}^{-1}$ based on the analysis of \citet[][see their Appendix A]{donnari:2019}.


\bsp	
\label{lastpage}
\end{document}